\documentclass[11pt,a4paper]{article}
\pdfoutput=1
\usepackage{jheppub}
\usepackage{amsmath}
\usepackage{dsfont}
\usepackage{MnSymbol}
\usepackage{ulem}
\usepackage{array}
\usepackage{multirow}
\usepackage{hhline}

\makeatletter\renewcommand{\@biblabel}[1]{#1.}\makeatother

\def\be{\begin{equation}}
\def\ee{\end{equation}}

\def\nn{\nonumber}
\def\rd{{\rm d}}
\def\bs{\boldsymbol}
\def\i{{\rm i}}
\def\e{{\rm e}}
\def\a{{\sf a}}
\def\V{{\sf V}}
\def\S{{\sf S}}
\def\Q{{\sf Q}}
\def\P{{\sf P}}
\def\T{{\sf T}}
\def\Z{{\sf Z}}
\def\ul#1{\underline{#1}}

\def\ket#1{{ |#1\rangle}}
\def\bra#1{{ \langle#1|}}
\def\braket #1#2{{ \langle #1|#2 \rangle}}

\preprint{
{\small{\textsf{}}}}

\title{$q$-Virasoro modular double\\
and 3d partition functions}

\author[a]{Anton Nedelin,}
\author[b]{Fabrizio Nieri,}
\author[b]{Maxim Zabzine.}

\affiliation[a]{Dipartimento di Fisica, Universit\`a di Milano-Bicocca,\\
Piazza della Scienza 3, I-20126 Milano, Italy,\\[3pt]
and\\[3pt]
INFN, sezione di Milano-Bicocca,\\
I-20126 Milano, Italy.}
\affiliation[b]{Department of Physics and Astronomy, Uppsala University,\\
Box 516, SE-75120 Uppsala, Sweden.}

\emailAdd{anton.nedelin@unimib.it}
\emailAdd{fb.nieri@gmail.com}
\emailAdd{maxim.zabzine@physics.uu.se}

\abstract{We study partition functions of 3d $\mathcal{N}=2$ ${\rm U}(N)$ gauge theories on compact manifolds which are $S^1$ fibrations over $S^2$. 
We show that the partition functions are free field correlators of vertex operators and screening charges of the \mbox{$q$-Virasoro} modular double, which we define. The inclusion of supersymmetric Wilson loops in arbitrary representations allows us to show that the generating functions of Wilson loop vacuum expectation values satisfy two ${\rm SL}(2,\mathbb{Z})$-related commuting sets of \mbox{$q$-Virasoro} constraints. We generalize our construction to 3d $\mathcal{N}=2$ unitary quiver gauge theories and as an example we give the free boson realization of the ABJ(M) model.}


\arxivnumber{1605.07029 [hep-th]}

\begin{document}

\maketitle
\flushbottom
%\newpage

\section{Introduction}
Quantum field theories in three dimensions have played a relevant role in theoretical physics and many branches of mathematics since Witten's seminal work on \mbox{Wilson loops} in Chern-Simons theory and Jones polynomials \cite{Witten:1988hf}. Chern-Simons theory and its refined \cite{Aganagic:2011sg,Aganagic:2012ne,Aganagic:2012hs} or supersymmetric extensions feature prominently in topological string theory \cite{Witten:1992fb,Gopakumar:1998ki,Aganagic:2002qg}, in the study of the low energy physics of string/M-theory through Hanany-Witten brane constructions \cite{Hanany:1996ie} or the celebrated ABJ(M) model for the effective theory of M2 branes \cite{Aharony:2008ug,Aharony:2008gk}.  The application of field theory methods for studying 3d manifolds and knot theory has recently produced many new results and connections between the two fields, culminated in the discovery of the 3d-3d correspondence \cite{Dimofte:2010tz,Dimofte:2011ju,Dimofte:2011py,Chung:2014qpa} (see also the review \cite{Dimofte:2014ija}) relating 3d $\mathcal{N}=2$ SCFTs arising from M5 branes compactified on a 3d manifold to complex Chern-Simons on the latter \cite{Witten1991}. More generally, the embedding of 3d $\mathcal{N}=2$ Chern-Simons-Yang-Mills theories in string/M-theory has provided many insights into their physics, including 3d Seiberg-like dualities \cite{Karch:1997ux,Aharony:1997gp,Niarchos:2008jb,Benini:2011mf,Aharony:2013dha,Aharony:2013kma} and mirror symmetry \cite{Intriligator:1996ex,deBoer:1997ka,Aharony:1997bx}. However, a better understanding of the rich dynamics and web of dualities of these field theories is desirable, perhaps exploiting some large symmetry hidden in this class of theories: this is the topic of this work.

In this paper we focus on a wide class of 3d $\mathcal{N}=2$ Yang-Mills-Chern-Simons (YM-CS) unitary quiver gauge theories. A simple yet instrumental example for our analysis is the ${\rm U}(N)$ theory coupled to 1 adjoint chiral multiplet, which has a distinguished role also within the 3d-3d correspondence \cite{Gadde:2013sca,Chung:2014qpa,Gukov:2015sna}. These theories can conveniently be studied on compact backgrounds \cite{Festuccia:2011ws,Dumitrescu:2012ha,Closset:2012ru,Closset:2013vra,Closset:2014uda,Imbimbo:2014pla} such as the squashed $S^3_b$, lens spaces $L(r,1)$ and $S^2\times S^1$.  The application of supersymmetric localization \cite{Pestun:2007rz} to this class of theories \cite{Kapustin:2009kz,Hama:2011ea,Imamura:2011wg,Alday:2013lba,Gang:2009wy,Benini:2011nc,Alday:2012au,Imamura:2012rq,Imamura:2013qxa,Imamura:2011su,Kapustin:2011jm,Benini:2015noa,Benini:2016hjo} has been a powerful tool for studying non-perturbative gauge dynamics over the past few years. In fact, one of the main outcome of the localization method is that expectation values of supersymmetric observables can be exactly computed by reducing path integrals to finite dimensional matrix models (Coulomb branch localization), which can be then analyzed from different angles. 
Our goal is to use these results to show that there is a universal algebraic structure underlying the supersymmetric sector of these theories in any such background, which we call the {\it $\textit{W}_{q,t}$  modular double}, or {\it \mbox{$q$-Virasoro} modular double} for the single node quiver.

In order to explain our results it is enough to consider the reference example given by the ${\rm U}(N)$ theory coupled to 1 adjoint chiral multiplet and possibly \mbox{(anti-)fundamental} chiral multiplets, in which case the Coulomb branch partition function of the theory can be schematically written as
\be
Z=\sum_{\ul{\ell}\in\mathbb{F}^N}\int\!\rd^N \ul{x}\;\Delta(\ul{x},\ul{\ell})\;\e^{\sum_jV(x_j,\ell_j)}~,\nn
\ee
where the continuous and discrete variables $\{\ul{x},\ul{\ell}\}=\{x_j,\ell_j,j=1,\ldots,N\}$ parametrize the localization locus, $\Delta(\ul{x},\ul{\ell})$ is the 1-loop contribution of the vector and adjoint multiplets, $\exp(\sum_j V(x_j,\ell_j))$ is the 1-loop contribution of (anti-)fundamental matter and classical CS action, while the sum is over the different topological sectors: \mbox{$\mathbb{F}=(\emptyset,\mathbb{Z}_r,\mathbb{Z})$} for $S^3_b$, $L(r,1)$ and $S^2\times S^1$ respectively. One of our main results is that 3d compact space partition functions are free boson correlators of vertex operators ($\mathcal{V}(z)$) and integrated screening currents ($\mathcal{S}(x)$) of a modular double version of the \mbox{$q$-Virasoro} algebra \cite{Shiraishi:1995rp} which we define, namely
\be
Z=\int\!\rd^N \ul{x}\; \langle\; \prod_{f} \mathcal{V}_f(z_f)\prod_{j=1}^N \mathcal{S}(x_j)\;\rangle~,\nn
\ee
leading to a dual 2d CFT-like description in the spirit of the AGT correspondence \cite{Alday:2009aq,Wyllard:2009hg,Awata:2009ur} and similar to the proposal of \cite{Nieri:2013yra,Nieri:2013vba} for the SQED. The origin of the $q$-deformation has been proposed to lie in the little string deformation of the 6d $(2,0)$ theory \cite{Aganagic:2015cta}. 

The central object of our construction is the modular double screening current $\mathcal{S}(x)$. We define it to be the operator which commutes, up to total differences, with two commuting copies ($i=1,2$) of the \mbox{$q$-Virasoro} generators $\{\T_{n,i},n\in\mathbb{Z}\}_{i=1,2}$ whose \mbox{$q$-deformation} parameters are related by ${\rm SL}(2,\mathbb{Z})$ transformations, namely
\be
[\T_{n,1},\T_{m,2}]=0~,\quad [\T_{n,i},\mathcal{S}(x)]=\textrm{total difference}~,\nn
\ee
\be
q_1=\e^{2\pi\i\epsilon}~,\quad q_2=\e^{-2\pi\i g\cdot \epsilon}~,\nn
\ee
where $g\cdot$ is the standard $g\in{\rm SL}(2,\mathbb{Z})$ action on the modular parameter $\epsilon$. In order to avoid possible confusion, throughout this paper the index $i=1,2$ will be exclusively used for distinguishing the two copies and nothing else. The $q_i$ are related to geometric moduli of the gauge theory background (squashing or fibration parameters), in fact
\be
g\cdot\epsilon=\frac{\epsilon}{1-r\epsilon}~\nn
\ee
for the lens space $L(r,1)$, whereas $r=1,0$ for the particular cases $S^3_b$ and $S^2\times S^1$ respectively. The fact that the $q$-deformation parameters are related by ${\rm SL}(2,\mathbb{Z})$ is crucial: in this case we can give $\mathcal{S}(x)$ in terms of the screening currents $\S(w)_i$ of the individual \mbox{$q$-Virasoro} copies according to
\be
\mathcal{S}(x)=\sum_{\ell\in\mathbb{F}}w(x,\ell)_1 \; w(x,\ell)_2\; \S(w(x,\ell))_1\otimes \S(w(x,\ell))_2~,\nn
\ee
where the dependence of the summands on the gauge theory continuous and discrete variables is through the ``holomorphic" coordinate $w(x,\ell)_1$ and its ``conjugate" in the sense of the ${\rm SL}(2,\mathbb{Z})$ pairing, which acts also on the position in a certain way
\be
w(x,\ell)_2=g\cdot w(x,\ell)_1~.\nn
\ee
From the gauge theory viewpoint, $w(x,\ell)_i$ are supersymmetric Wilson lines ($\bs w_i$) at the North ($i=1$) and South ($i=2$) poles of the $S^2$ base
\be 
\bs w_i=\textrm{Pexp}\left(\i\oint_{\mathcal{C}_i}(A-\i\sigma|\dot{\mathcal{C}_i}|\rd s)\right)~,\nn
\ee
evaluated at the localization locus and projected on a ${\rm U}(N)$ fundamental weight $\rho$
\be
w(x,\ell)_1=\rho\left(\bs w(\ul{x},\ul{\ell})_1\right)~,\quad w(x,\ell)_2=\rho\left(\bs w(\ul{x},\ul{\ell})_2\right)~,\nn
\ee
whose expressions depend on the specific background. Here $A$ denotes the gauge connection, $\sigma$ the vector multiplet scalar and $\mathcal{C}_i$ a supersymmetry preserving cycle.

The partition function is not the most general observable one can consider. Following what we have just mentioned, an important class of observables which can be computed through localization is given by supersymmetric Wilson loops. The evaluation of Wilson loop vacuum expectation values (v.e.v.) at the North or South poles of the $S^2$ base of the geometries we are considering amounts to insert
\be
\textrm{Tr}_{\mathcal{R}_i}\left(\bs w(\ul{x},\ul{\ell})_i\right)\nn
\ee
into the Coulomb branch partition function, where $\mathcal{R}_i$ is a representation of the ${\rm U}(N)$ gauge group. Using the standard character decomposition, we can package Wilson loop v.e.v.'s in arbitrary representations into the generating function
\be
Z(\ul \tau_1,\ul \tau_2)=\sum_{\mathcal{R}_1,\mathcal{R}_2}\sum_{\ul{\ell}\in\mathbb{F}^N} \!\int\!\rd^N \ul x\;\Delta(\ul{x},\ul{\ell})\;\e^{\sum_jV(x_j,\ell_j)}\!\prod_{i=1,2} \textrm{Tr}_{\mathcal{R}_i}(\bs v(\ul\tau_i))\textrm{Tr}_{\mathcal{R}_i}(\bs w(\ul{x},\ul{\ell})_i)~,\nn
\ee
where the additional insertions $\textrm{Tr}_{\mathcal{R}_i}(\bs v(\ul\tau_i))$ can be though of as background Wilson loops. The generating function is the natural object to consider from a matrix model perspective, and we can give it a \mbox{$q$-Virasoro} interpretation as well: for YM theories it can be identified with the heighest weight state
\be
Z(\ul \tau_1,\ul \tau_2)\simeq \int\!\rd^N \ul{x}\; \prod_{j=1}^N \mathcal{S}(x_j)\ket{\alpha}~.\nn
\ee
In this language, (anti-)fundamental matter can be coupled to the gauge theory by shifting the ``time" variables $\ul\tau_{i}$ (isomorphic to the creation operators in the free boson representation of \mbox{$q$-Virasoro}), or equivalently by acting on the state with additional vertex operators. Inclusion of CS terms can be dealt with similarly. Remarkably, our identification implies the existence of two ${\rm SL}(2,\mathbb{Z})$-related commuting sets of \mbox{$q$-Virasoro} constraints (or Ward identities) satisfied by the YM generating function 
\be
T_n(\ul \tau_i)Z(\ul \tau_1,\ul \tau_2)=0~,\quad n>0~,\nn
\ee
where $T_n(\ul \tau_i)\simeq \T_{n,i}$ are differential operators in $\ul \tau_i$, which express  the highest weight condition of the YM generating function. A similar description holds when including CS terms. This observation opens up the possibility of characterizing compact space generating functions as solutions of two infinite sets of PDEs. Similar considerations have been put forward in \cite{Dimofte:2011ju,Dimofte:2011py,Aganagic:2011mi,Gadde:2013wq,Chung:2014qpa}, where it is shown that the algebra of line operators and their action on 3d partition functions gives rise to recurrence relations quantizing classical spectral curves or knot polynomials, and in \cite{Gaiotto:2013bwa,Bullimore:2014awa,Koroteev:2015dja,Koroteev:2016znb}, where the relation of line operators with difference operators/quantum Hamiltonians of integrable systems is discussed. 

Our results fit nicely with the observed factorization properties of 3d compact space partition functions \cite{Pasquetti:2011fj,Dimofte:2011py,Beem:2012mb,Benini:2013yva,Fujitsuka:2013fga,Taki:2013opa,Bullimore:2014awa,Hwang:2015wna,Nieri:2015yia}. All the manifolds we are interested in admit indeed a decomposition into a pair of solid tori $(D^2\times  S^1)_{i=1,2}$, where the boundary homeomorphism is implemented by the $g\in {\rm SL}(2,\mathbb{Z})$ element acting on one boundary torus with modulus $\epsilon$. Gauge theory partition functions on the diverse compact spaces can be recovered by ${\rm SL}(2,\mathbb{Z})$ gluings of partition functions on the solid torus $D^2\times S^1$. Supersymmetric partition functions on such elementary background are known as 3d holomorphic blocks \cite{Dimofte:2011py}
\be
\mathcal{B}^{\rm 3d}_c=\oint_{c}\rd^N \ul{w}\; \Upsilon^{\rm 3d}(\ul{w})~,\nn
\ee
and for ${\rm U}(N)$ YM theories they have been shown \cite{Aganagic:2013tta} (see also the review \cite{Aganagic:2014kja}) to be captured  by free boson correlators of \mbox{$q$-Virasoro} screening currents and vertex operators (${\sf H}(z)$) 
\be
\mathcal{B}^{\rm 3d}_c=\oint_{c}\rd^N\ul{w}\; \langle\; \prod_{f} {\sf H}_f(z_f)\prod_{j=1}^N \S(w_j)\;\rangle~.\nn
\ee
Our construction then reveals the algebraic structure behind the observed non-trivial decomposition \cite{Beem:2012mb,Nieri:2015yia}
\be
Z=\sum_{\ul{\ell}\in\mathbb{F}^N}\int\!\rd^N \ul x\; \Upsilon^{\rm 3d}(\ul w(\ul x,\ul \ell))_1\Upsilon^{\rm 3d}(\ul w(\ul x,\ul \ell))_2~\nn
\ee
of compact space partition functions. Moreover, in many cases it has been shown that $Z$ can be completely factorized as
\be
Z= \sum_{\{c\}}\left(\mathcal{B}^{\rm 3d}_c\right)_1\left(\mathcal{B}^{\rm 3d}_c\right)_2~,\nn
\ee
where the sum is over the supersymmetric massive vacua of the effective 2d theory on the cigar or flat connections in complex CS through the 3d-3d correspondence. Our general results explain this property from the existence of two commuting sets of \mbox{$q$-Virasoro} constraints satisfied by the generating functions. 

Finally, our results can also be read in the context of the BPS/CFT correspondence and 5d AGT. Supersymmetric 5d unitary quiver gauge theories in the $\Omega$-background have an interesting class of observables known as $qq$-characters which have been recently constructed in \cite{Nekrasov:2015wsu} (building on previous works \cite{Nekrasov:2012xe,Nekrasov:2013xda}). In particular, it is shown in \cite{Kimura:2015rgi} that the $qq$-characters generate quiver ${\rm W}_{q,t}$ symmetry algebras and Ward identities for 5d (extended) Nekrasov partition functions \cite{Nekrasov:2002qd,Nekrasov:2003rj}. When 5d gauge theories can be engineered by M-theory compactifications on toric Calabi-Yau 3-folds \cite{Leung:1997tw,Kol:1998cf} or type IIB $(p,q)$-webs \cite{Aharony:1997ju,Aharony:1997bh}, one can also use the refined topological vertex formalism \cite{Iqbal:2007ii,Awata:2005fa,Taki:2007dh} to conveniently compute the 5d Nekrasov partition functions. Using this approach it has been recently realized \cite{Mironov:2016yue,Awata:2016riz} that 5d gauge theories supported on $(p,q)$-webs form a representation of the Ding-Iohara-Miki algebra \cite{1996qDH,2007JMPMIKI}, which is the building block for constructing ${\rm W}_{q,t}$ algebras (at least in the $A_n$ case) as much as the strip geometry \cite{Iqbal:2004ne} is the building block for constructing toric webs. In any case, the free boson representation of the relevant symmetry algebra yields a matrix model description of the 5d Nekrasov partition function \cite{Awata:2010yy,Mironov:2011dk,Carlsson:2013jka,Zenkevich:2014lca,Zenkevich:2015rua,Morozov:2015xya,Mironov:2015thk,Mironov:2016cyq}, which at isolated points on the Coulomb branch describes a 3d vortex theory \cite{Aganagic:2013tta,Aganagic:2014kja,Aganagic:2014oia,Aganagic:2015cta}. We thus expect that our construction describes the compact space version, or the non-perturbative completion in the sense of \cite{Lockhart:2012vp}, of this chain of dualities between gauge/string theory and quantum algebras.
 
The rest of this paper is organized as follows. In section \ref{sec:free_field} we review some basics in the theory of conformal matrix models and Virasoro constraints, and the analogous constructions for the $q$-deformed case. We also review the \mbox{$q$-Virasoro} description of 3d holomorphic blocks of ${\rm U}(N)$ YM theories, and propose the \mbox{$q$-Virasoro} interpretation of the Wilson loop generating function. In section \ref{sec:S3} we discuss in detail ${\rm U}(N)$ YM theories on the squashed $S^3_b$, focusing on the Coulomb branch partition function and Wilson loop generating function. We then discuss how these objects can be mapped to correlators or highest weight states of the \mbox{$q$-Virasoro} modular double and we find two ${\rm SL}(2,\mathbb{Z})$-related commuting sets of \mbox{$q$-Virasoro} constraints annihilating the generating function. We also comment on few interesting limits of the gauge theory and the associated constraints, such as the round $S^3$ or special values of the adjoint mass. In section \ref{sec:other} we extend our analysis to the lens space partition function, the index and twisted index. In section \ref{sec:CS} we discuss how CS terms can be described in the \mbox{$q$-Virasoro} side, leading to ``dressed" correlators and modified \mbox{$q$-Virasoro} constraints. In section \ref{sec:generalizations} we consider the generalization to quiver gauge theories, with special focus on the ABJ(M) theory, and the relation to quiver ${\rm W}_{q,t}$ algebras. In section \ref{sec:summary} we summarize our results and comment on open questions and interesting directions for future work, such as the possible 4d/elliptic lift of our construction and the relation to 5d theories. 

\section{Matrix models, free fields and 3d gauge theories}\label{sec:free_field}
In this section we summarize basic facts about the $\beta$-ensemble and its free boson Virasoro construction, while for a detailed review we refer to \cite{Morozov:1994hh,DiFrancesco:1993cyw}. This elementary discussion will allow us to introduce the main tools which also apply to the $q$-deformed $\beta$-ensemble and \mbox{$q$-Virasoro} algebra, for details we refer to \cite{Odake:1999un,Awata:2010yy}. We then recall the \mbox{$q$-Virasoro} interpretation of 3d partition functions on $D^2\times S^1$ given by \cite{Aganagic:2013tta}, and we extend the duality by mapping the Wilson loop generating function to the $q$-deformed $\beta$-ensemble.

\subsection{Virasoro matrix model}
Let us consider the matrix model 
\be
Z(\ul \tau)=\mathcal{N}_0\int\!\rd^N \!\ul{w}\; \Delta_\beta(\ul{w})\;\e^{\sqrt{\beta}\sum_j V(w_j|\ul \tau)}~,\quad V(w|\ul \tau)=\sum_{n>0}\tau_n w^n~,
\ee
where $\Delta_\beta(\ul{w})$ is the integration measure describing the interactions between the eigenvalues $\ul w$, $V(w|\ul \tau)$ is the potential whose shape is described by the time parameters $\ul \tau$, $\mathcal{N}_0$ is a normalization parametrized by $\tau_0$ and $\beta\in\mathbb{C}$. We refer to such a partition function as a Virasoro matrix model if it satisfies the Virasoro constraints
\be\label{virconstraints}
L_n(\ul \tau)Z(\ul \tau)=0~,\quad n\in\mathbb{Z}_{>0}~,
\ee
where $L_n(\ul \tau)$ are differential operators in the time variables satisfying the positive mode subalgebra of the full Virasoro algebra
\be\label{viralg}
[L_n,L_m]=(n-m)L_{n+m}+\frac{c}{12}n(n^2-1)\delta_{n+m,0}~,\quad n,m\in\mathbb{Z}~.
\ee
The fact that the matrix model partition function depends on infinitely many parameters and that it is subject to infinitely many constraints forming a closed algebra, is a strong indication that the matrix model can be defined as the (unique) solution to the Virasoro constraint equations (with suitable boundary conditions).

A simple method to built the matrix model satisfying Virasoro constraints is exploiting the free boson realization of the Virasoro algebra. Let us consider the Heisenberg algebra (we display non-trivial relations only)
\be
[\a_n,\a_m]=2n\delta_{n+m,0}~,\quad [\P,\Q]=2~, \quad n,m\in\mathbb{Z}\backslash \{0\}~,
\ee
and the Fock module $\mathcal{F}_\alpha$ over the charged vacuum $\ket{\alpha}$  spanned by the states
\be
\mathcal{F}_\alpha=\left\{\prod_{n=1}^{|\mu|}\a_{-\mu_n}\ket{\alpha}\right\}~, \quad \prod_{n=1}^{|\mu|}\a_{\mu_n}\ket{\alpha}=0~,\quad \ket{\alpha}=\e^{\frac{\alpha}{2}\Q}\ket{0}~, \quad \P\ket{\alpha}=\alpha\ket{\alpha}~,
\ee 
for any partition $\mu$ of length $|\mu|$ and given momentum $\alpha\in\mathbb{C}$. The operators
\be\label{Virgen}
\begin{split}
{\sf L}_n&=\frac{1}{4}\sum_{k\neq 0,n } :\a_{n-k}\a_k:+\frac{1}{2}\a_n\P-\frac{1}{2}Q_\beta(n+1)\a_n~,\quad n\neq 0~,\\
{\sf L}_0&=\frac{1}{2}\sum_{k>0}\a_{-k}\a_k+\frac{\P^2}{4}-\frac{1}{2}\P Q_\beta~,\quad Q_\beta=\sqrt{\beta}-\frac{1}{\sqrt{\beta}}~,
\end{split}
\ee
where $:~:$ denotes normal ordering (i.e. positive modes to the right of negative modes and $\P$ to the right of $\Q$), close the Virasoro algebra (\ref{viralg}) with central charge $c=1-6Q_\beta^2$. Using the algebra representation 
\be\label{viriso}
\a_{-n}\simeq n \tau_n~,\quad \a_n\simeq2\frac{\partial}{\partial \tau_n}~,\quad \Q \simeq \tau_0~,\quad \P\simeq 2\frac{\partial}{\partial \tau_0}~,\quad \ket{\alpha}=\e^{\frac{\alpha}{2}\Q}\ket{0}\simeq \e^{\tau_0\frac{\alpha}{2}}\cdot 1~,
\ee
we get the differential representation 
\be\label{Lop}
{\sf L}_n\simeq L_n(\ul \tau)=\sum_{k\geq 0}k \tau_k\frac{\partial}{\partial \tau_{n+k}}+\sum_{k=0}^{n}\frac{\partial^2}{\partial \tau_{n-k}\partial \tau_k}-Q_\beta(n+1)\frac{\partial}{\partial \tau_n}~,\quad n>0~.
\ee
Then the original problem (\ref{virconstraints}) can be solved by finding a free boson operator $\S(w)$ whose commutator with the Virasoro generators is a total derivative, namely
\be\label{virscreening}
[{\sf L}_n,\S(w)]=\frac{\rd}{\rd w}{\sf O}(w)~
\ee
for some ($n$-dependent) operator ${\sf O}(w)$. In fact, by defining\footnote{A suitable choice of integration contour is to be understood.}
\be
\Z={\sf J}^N~,\quad {\sf J}=\int\!\rd w\; \S(w)~,
\ee
through the representation (\ref{viriso}) we immediately get\footnote{This identification is usually achieved by an explicit projection onto the coherent state $\bra{\alpha_\infty}{\sf G}(\ul\tau)$, ${\sf G}(\ul\tau)=\exp(\frac{1}{2}\sum_{n>0}\tau_n\a_n)$, with $\alpha_\infty=\alpha+2\sqrt{\beta}N$ and $\bra{\alpha_\infty}$ the dual charged Fock vacuum ($\braket{\alpha_a}{\alpha_b}=\delta_{ab}$). This projection is equivalent to the representation (\ref{viriso}): $2\frac{\partial}{\partial \tau_n}\bra{\alpha_\infty}{\sf G}(\ul\tau)=\bra{\alpha_\infty}{\sf G}(\ul\tau)\a_{n}$, $n\tau_n\bra{\alpha_\infty}{\sf G}(\ul\tau)=\bra{\alpha_\infty}{\sf G}(\ul\tau)\a_{-n}$, $\alpha_\infty\bra{\alpha_\infty}{\sf G}(\ul\tau)=\bra{\alpha_\infty}{\sf G}(\ul\tau)\P=2\frac{\partial}{\partial \tau_0}\bra{\alpha_\infty}{\sf G}(\ul\tau)$.}
\be
\Z\ket{\alpha}\simeq Z(\ul \tau)~,\quad {\sf L}_n\Z\ket{\alpha}\simeq L_n(\ul \tau)Z(\ul \tau)=0~,\quad n>0~,
\ee
where the last equality follows from the screening charge conservation $[{\sf L}_n,{\sf J}]=0$ and the highest weight condition ${\sf L}_{n>0}\ket{\alpha}=0$.\footnote{For $\alpha=0$ there are the additional constraints $n=-1,0$ due to $\mathfrak{sl}_2$ invariance of the vacuum $\ket{0}$.} Notice that by packaging the Virasoro generators into the current (stress tensor) ${\sf L}(z)=\sum_{n\in\mathbb{Z}}{\sf L}_{n}z^{-n-2}$, the constraints (\ref{virconstraints}) are equivalent to the regularity condition (Ward identities)
\be
z^2 {L}(z|\ul \tau)Z(\ul \tau)={\rm Pol}(z)~,
\ee 
for a certain ($\ul \tau$-dependent) polynomial ${\rm Pol}(z)$.

The central object of the free boson construction is the screening current $\S(w)$ defined by (\ref{virscreening}), and its free boson representation is given by
\be
\S(w)=\; :\e^{-\sqrt{\beta}\sum_{n\neq 0}\frac{w^{-n}}{n}\a_n}:\e^{\sqrt{\beta}\Q}w^{\sqrt{\beta}\P}~.
\ee
This representation allows us to write down the matrix model explicitly, in fact
\be\label{rVan}
\prod_{j=1}^N \S(w_j)=\; :\prod_{j=1}^N \S(w_j):\; \Delta_\beta(\ul{w})~,\quad \Delta_\beta(\ul{w})=\prod_{k<j}(w_k-w_j)^{2\beta}~,
\ee
and hence
\be
\Z\ket{\alpha}=\int\!\rd^N \ul{w}\; \Delta_\beta(\ul w)\prod_j w_j^{\sqrt{\beta}\alpha}\e^{\sqrt{\beta}\sum_{n>0}\sum_j \frac{w_j^n}{n}\a_{-n}}\e^{\sqrt{\beta}N\Q}\ket{\alpha}\simeq Z(\ul \tau)~,
\ee
with $V(w|\ul \tau)= \alpha\ln w+\sum_{n> 0}\tau_n w^n$, $\mathcal{N}_0=\exp\sqrt{\beta}\tau_0(N+\frac{\alpha}{2\sqrt{\beta}})$, where we used (\ref{viriso}).

We conclude this brief review of Virasoro matrix models with three remarks. Firstly, this 2d CFT construction can be easily generalized to other (quantum) algebras provided that the free boson representation of generators and screening currents is known. For instance, this is the case with ${\rm W}$ algebras and their $q$-deformation, which will be in fact the main focus of this paper. Secondly, different looking matrix models may actually be related by a simple redefinition of the time variables. Thirdly, the matrix model can be enriched through the inclusion of vertex operators in the free boson correlator. In 2d CFTs there is a distinguished set of operators called primaries, which in the free boson representation are given by
\be\label{virvertex}
{\sf H}_\gamma(z)=\; :\e^{-\frac{\gamma}{2}\sum_{n\neq 0}\frac{z^{-n}}{n}\a_n}:\e^{\frac{\gamma}{2}\Q}z^{\frac{\gamma}{2}\P}~,\quad \gamma\in\mathbb{C}~.
\ee
Their OPE with the screening current is
\be
{\sf H}_\gamma(z)\S(w)=\;:{\sf H}_\gamma(z)\S(w):(1-w z^{-1})^{\sqrt{\beta}\gamma}z^{\sqrt{\beta}\gamma}~,
\ee
whose effect is simply to add a constant background to the time variables
\be
\tau_n\to \tau_n-\gamma\frac{z^{-n}}{n}~,
\ee
and to change the normalization by a constant multiplicative factor. For this reason we will be mainly interested in theories without vertex operators.

\subsection{$q$-Virasoro matrix model}
The \mbox{$q$-Virasoro} algebra is the associative algebra generated by $\{T_n,~n\in\mathbb{Z}\}$ satisfying the relation \cite{Shiraishi:1995rp}
\be
f\left(\frac{w}{z}\right)T(z)T(w)-f\left(\frac{z}{w}\right)T(w)T(z)=-\frac{(1-q)(1-t^{-1})}{(1-p)}\left(\delta\left(p \frac{w}{z}\right)-\delta\left(p^{-1}\frac{w}{z}\right)\right)~,
\ee
where
\be
T(z)=\sum_{n\in\mathbb{Z}}T_n z^{-n}~,\quad f(z)=\sum_{\ell\geq 0}f_\ell z^\ell =\e^{\sum_{n>0}\frac{(1-q^n)(1-t^{-n})}{n(1+p^{n})}z^n}~,\quad \delta(z)=\sum_{n\in\mathbb{Z}}z^n~,
\ee
and $q,t,p\in\mathbb{C}$ with $p=qt^{-1}$. This algebra provides a 1-parameter deformation of the Virasoro algebra. In fact, upon setting $t=q^{\beta}$, $q=\e^{\hbar}$, we have the small $\hbar\in\mathbb{R}$ expansion 
\be\label{virlimit}
T_n=2\delta_{n,0}+\hbar^2 \beta \left(L_n+\frac{Q_\beta^2}{4}\delta_{n,0}\right)+O(\hbar^4)~,
\ee
where the operators $L_n$ close the Virasoro algebra (\ref{viralg}) with central charge $c=1-6 Q_\beta^2$.

The Heisenberg algebra (we display non-trivial relations only)
\be\label{qosc}
\!\!\!\![\a_n,\a_m]=\frac{1}{n}(q^\frac{n}{2}-q^{-\frac{n}{2}})(t^\frac{n}{2}-t^{-\frac{n}{2}})(p^\frac{n}{2}+p^{-\frac{n}{2}})\delta_{n+m,0}~, \quad [\P,\Q]=2~,\quad n,m\in\mathbb{Z}\backslash \{0\}~,
\ee
gives a free boson representation of the \mbox{$q$-Virasoro} algebra according to
\be
\T(z)=\sum_{n\in\mathbb{Z}}\T_n z^{-n}=\sum_{\sigma=\pm 1}\Lambda_\sigma(z)~,\quad \Lambda_\sigma(z)=\; :\e^{\sigma\sum_{n\neq 0}\frac{z^{-n}}{(1+p^{-\sigma n})}\a_n}:q^{\sigma\frac{\sqrt{\beta}}{2}\P}p^{\frac{\sigma}{2}}~,
\ee
where $\beta=\ln t/\ln q$. The \mbox{$q$-Virasoro} screening current is given by\footnote{There is another screening current with $q\to t^{-1}$, $\sqrt{\beta}\to-\frac{1}{\sqrt{\beta}}$, but we do not need it in this work.}
\be\label{SqVir}
\S(w)=\; :\e^{-\sum_{n\neq 0}\frac{w^{-n}}{q^{n/2}-q^{-n/2}}\a_n}:\e^{\sqrt{\beta}\Q}w^{\sqrt{\beta}\P}~.
\ee
One can indeed verify that the defining relation
\be\label{TS}
[\T_n,\S(w)]=\frac{{\sf O}(q w)-{\sf O}(w)}{w}~
\ee
holds true for a certain ($n$-dependent) operator ${\sf O}(w)$, implying the conservation of the screening charge for a suitable contour, $[\T_n,\oint \!\rd w\; \S(w)]=0$.

The \mbox{$q$-Virasoro} matrix model can now be constructed by exploiting the strategy outlined in the previous subsection. The product of several \mbox{$q$-Virasoro} screening currents yields 
\be
\prod_{j=1}^N \S(w_j)=\; :\prod_{j=1}^N \S(w_j):\; \Delta_\beta(\ul w;q)c_\beta(\ul w,1;q)\prod_j w_j^{\beta(N-1)}~,
\ee
with
\be\label{Deltabeta}
\Delta_\beta(\ul w;q)=\prod_{k\neq j}\frac{(w_k w_j^{-1};q)_\infty}{(t w_k w_j^{-1};q)_\infty}~,\quad c_\beta(\ul w,m;q)=\prod_{k<j}(w_k w_j^{-1})^\beta\frac{\Theta(t m w_k w_j^{-1};q)}{\Theta(m w_k w_j^{-1};q)}~, 
\ee
or
\be\label{Deltabeta2}
\Delta_\beta(\ul w;q)c_\beta(\ul w;q)=\prod_{k<j}(w_k w_j^{-1})^\beta(1-w_j w_k^{-1})\frac{(q t^{-1} w_j w_k^{-1};q)_\infty}{(t w_j w_k^{-1};q)_\infty}~,
\ee
where the $q$-Pochhammer symbol and $\Theta$ function are defined in (\ref{qPoch}) and (\ref{Theta}) respectively. As before, we can now define the operator
\be
\Z={\sf J}^N~,\quad {\sf J}=\oint\!\rd^N w\;\S(w)~,
\ee
and consider the state
\begin{align}
\Z\ket{\alpha}&=\oint\!\frac{\rd^N \ul w}{2\pi\i\ul w}\; \Delta_\beta(\ul w;q)c_\beta(\ul w,1;q)\prod_j w_j^{\sqrt{\beta}(\alpha+\sqrt{\beta}N-Q_\beta)}\e^{\sum_{n>0}\frac{\sum_j w_j^n}{q^{n/2}-q^{-n/2}}\;\a_{-n}}\e^{\sqrt{\beta}N\Q}\ket{\alpha}~.
\end{align}
Finally, the algebra representation 
\be\label{qviriso}
\begin{split}
\a_{-n}&\simeq(q^\frac{n}{2}-q^{-\frac{n}{2}}) \tau_n~,\quad \a_n\simeq\frac{1}{n}(t^{\frac{n}{2}}-t^{-\frac{n}{2}})(p^{\frac{n}{2}}+p^{-\frac{n}{2}})\frac{\partial}{\partial \tau_n}~,\quad n\in\mathbb{Z}_{>0}~,\\
\sqrt{\beta}\Q&\simeq \tau_0~,\quad \P\simeq\frac{2}{\sqrt{\beta}}\frac{\partial}{\partial \tau_0}~,\quad\ket{\alpha}=\e^{\frac{\alpha}{2}\Q}\ket{0}\simeq \e^{\tau_0\frac{\alpha}{2\sqrt{\beta}}}\cdot 1~,
\end{split}
\ee
yields the matrix model 
\be\label{qvirgen}
\begin{split}
\Z\ket{\alpha}\simeq Z(\ul \tau)&=\mathcal{N}_0\oint\!\frac{\rd^N \ul w}{2\pi\i\ul w}\; \Delta_\beta(\ul w;q)c_\beta(\ul w,1;q)\e^{\sum_j V(w_j|\ul \tau)}~,\\
V(w|\ul \tau)&=\sqrt{\beta}(\alpha+\sqrt{\beta}N-Q_\beta)\ln w+\sum_{n>0}\tau_n w^n~,\quad \mathcal{N}_0=\e^{\tau_0(N+\frac{\alpha}{2\sqrt{\beta}})}~.
\end{split}
\ee
Due to the conservation of the screening charge $[\T_n,{\sf J}]\!=\!0$ and the highest weight condition $\T_{\!\!n>0}\ket{\alpha}\!=\!0$, the above partition function satisfies \mbox{$q$-Virasoro} constraints by construction \cite{Awata:2010yy}
\be\label{qvirconstraints}
T(z|\ul\tau)Z(\ul \tau)={\rm Pol}(z) \quad \Rightarrow\quad T_n(\ul \tau) Z(\ul \tau)=0~,\quad n>0~,
\ee
where the operators $T_n(\ul \tau)$ can be read from the modes $\T_n$ using (\ref{qviriso}). Explicitly, we have
\be\label{TnBell}
\T_n=\left\{\begin{array}{ll}\sum_{\sigma=\pm 1}q^{\sigma\frac{\sqrt{\beta}}{2}\P}p^\frac{\sigma}{2}\sum_{k\geq 0}\frac{B_k(\{A^{(\sigma)}_{-k}\})B_{n+k}(\{A^{(\sigma)}_{n+k}\})}{(n+k)!k!}~,\quad &\quad n\geq 0\\
\sum_{\sigma=\pm 1}q^{\sigma\frac{\sqrt{\beta}}{2}\P}p^\frac{\sigma}{2}\sum_{k\geq 0}\frac{B_{k-n}(\{A^{(\sigma)}_{n-k}\}) B_k(\{A^{(\sigma)}_{k}\})}{(k-n)!k!}~,\quad &\quad n<0 
\end{array}\right.~,
\ee
where we set 
\be
A^{(\sigma)}_n=\sigma\frac{\a_n |n|!}{(1+p^{-\sigma n})}~, \quad B_n(\{A_n\})=B_n(A_1,\ldots, A_n)~,\quad B_0=1~,
\ee
with $B_n(\{A_n\})$ the complete Bell polynomial defined by $\exp{\sum_{n>0}\frac{A_n}{n!}z^n}=\sum_{n\geq 0} \frac{B_n(\{A_n\})}{n!}z^n$.

As we have remarked at the end of the previous subsection, we can enrich the matrix model with the inclusion of vertex operators. The $q$-deformation of the operator (\ref{virvertex}) which is usually employed is \cite{Awata:2010yy} (see also \cite{1751-8121-49-34-345201} for a recent discussion)
\be\label{Hvertex}
{\sf H}_\gamma(z)=\; :\e^{-\sum_{n\neq 0}\frac{(t^{\gamma n/2}-t^{-\gamma n/2})z^{-n}}{(q^{n/2}-q^{-n/2})(t^{n/2}-t^{-n/2})}\;\lambda_n}:\e^{\frac{\gamma}{2}\sqrt{\beta}\Q}z^{\frac{\gamma}{2}\sqrt{\beta}\P}~,
\ee
where we have introduced the new basis 
\be\label{lambdabasis}
\lambda_{n}=\a_{n}(p^{n/2}+p^{-n/2})^{-1}~
\ee 
of the Heisenberg algebra. The OPE with the \mbox{$q$-Virasoro} screening current is
\be
{\sf H}_\gamma(z)\S(w)=\; : {\sf H}_\gamma(z)\S(w):\frac{(q^{\frac{1}{2}}t^{-\frac{\gamma}{2}}w z^{-1};q)_\infty}{(q^{\frac{1}{2}}t^{\frac{\gamma}{2}}w z^{-1};q)_\infty}z^{\beta\gamma}~,
\ee
and thus the inclusion of such a vertex operator simply amounts to add a constant background to the time variables
\be\label{qtaushift}
\tau_n\to\tau_n -\frac{(q^{\frac{1}{2}}t^{-\frac{\gamma}{2}}z^{-1})^n}{n(1-q^n)}+\frac{(q^{\frac{1}{2}}t^{\frac{\gamma}{2}}z^{-1})^n}{n(1-q^n)}~,
\ee
and to modify the normalization by a constant multiplicative factor. However, for our purposes it is also convenient to consider the ``half" vertex operator
\be\label{Vvertex}
\V_\gamma(z)=\; :\e^{-\sum_{n\neq 0}\frac{t^{\gamma n/2}z^{-n}}{(q^{n/2}-q^{-n/2})(t^{n/2}-t^{-n/2})}\;\lambda_n}:\e^{\frac{\gamma}{4}\sqrt{\beta}\Q}z^{\frac{\gamma}{4}\sqrt{\beta}\P}~,
\ee
whose inclusion will shift the time variables by the last term in (\ref{qtaushift}). With this definition, we also have
\be\label{Hsplitting}
{\sf H}_\gamma(z)=\; :\V_\gamma(z)\V_{-\gamma}(z)^{-1}:~.
\ee

\subsection{Gauge theory on $D^2\times S^1$}\label{D2xS1}
In this subsection we review a relevant application of the \mbox{$q$-Virasoro} theory for the study of 3d $\mathcal{N}=2$ gauge theories, and we also propose a gauge theory interpretation of the \mbox{$q$-Virasoro} matrix model (\ref{qvirgen}), extending the results of \cite{Aganagic:2013tta,Aganagic:2014kja}.

Partition functions of 3d $\mathcal{N}=2$  gauge theories compactified on $D^2\times S^1$ are computed by 3d holomorphic block integrals  introduced in \cite{Beem:2012mb} (see also \cite{Yoshida:2014ssa} for derivation through localization)
\be
\mathcal{B}^{\rm 3d}_c=\oint_{c}\frac{\rd^{{\rm rk }G}\ul w}{2\pi\i \ul w}\;\Upsilon^{\rm 3d}(\ul w)~,\nn
\ee
where the integral kernel $\Upsilon^{\rm 3d}(\ul w)$ is determined by the specific theory with gauge group $G$ and the integration is over a basis of middle dimensional cycles $\{c=1,\ldots\}$ in $(\mathbb{C}^\times)^{\textrm{rk}G}$. The vector multiplet contributes with the 1-loop factor\footnote{There can be anomalous terms in the 1-loop factors represented by quadratic polynomials. We will simply omit these factors because they vanish for the theories we are considering in this section.}
\be
\Upsilon_{\rm vec}^{\rm 3d}(\ul w)=
\prod_{\alpha\neq 0}(w_\alpha;q)_\infty~,
\ee
where $\alpha$ is a root of the gauge Lie algebra, while a chiral multiplet in a gauge representation $\mathcal{R}$ contributes with
\be
\Upsilon_{\rm N}^{\rm 3d}(\ul w, m)=\prod_{\rho\in\mathcal{R}}\frac{1}{(w_\rho m;q)_\infty}~,\quad\textrm{ or }\quad
\Upsilon_{\rm D}^{\rm 3d}(\ul w, m)=\prod_{\rho\in\mathcal{R}}(q w_\rho^{-1} m^{-1};q)_\infty~,
\ee
where $\rho$ is a weight of $\mathcal{R}$, $m$ is a global ${\rm U}(1)$ fugacity and the index N or D refers to Neumann or Dirichlet boundary conditions. The $\epsilon$ parameter 
\be
q=\e^{2\pi\i\epsilon}
\ee
can be interpreted as the disk equivariant parameter ($D^2\times S^1\simeq \mathbb{R}^2_{\epsilon}\times S^1$) or as the modular parameter of the boundary torus ($\partial( D^2\times S^1)\simeq \mathbb{T}^2$). Moreover, one can consider additional 2d vector, Fermi or chiral multiplets on the boundary torus and contributing to the integral kernel respectively with \cite{Benini:2013nda,Benini:2013xpa,Gadde:2013ftv}
\be\label{Theta1loop}
\prod_{\alpha\neq 0}\Theta(w_\alpha;q)~,\quad \prod_{\rho\in\mathcal{R}}\Theta(w_\rho m;q)^{\pm 1}~.
\ee
These contributions are important to introduce the correct CS units and ensure local and large gauge invariance \cite{Yoshida:2014ssa,Beem:2012mb,Gadde:2013wq}.

As a concrete example we can consider the ${\rm U}(N)$ YM theory coupled to 1 adjoint chiral multiplet and additional $N_{\rm f}$ fundamental and anti-fundamental chiral multiplets, in which case the 3d holomorphic block integral can be written as
\be
\mathcal{B}^{\rm 3d}_c=\oint_c\frac{\rd^N \ul w}{2\pi\i \ul w}\prod_{k\neq j=1}^N\frac{(w_k w_j^{-1};q)_\infty}{(m_{\rm a} w_k w_j^{-1};q)_\infty}\prod_{j=1}^N w_j^{\kappa_1}\prod_{f=1}^{N_{\rm f}}\frac{(q w_j \bar m_f;q)_\infty}{(w_j m_f;q)_\infty}~,
\ee
where $m_{\rm a}$ is the adjoint fugacity, $m_f,\bar m_f$ are the fugacities for the fundamental and anti-fundamentals and we also turned on the Fayet-Iliopoulos (FI) parameter $\kappa_1$. Notice that the multiplet content is proper of an $\mathcal{N}=4$ theory, namely the vector and adjoint chiral multiplets form a $\mathcal{N}=4$ vector multiplet, while the fundamental and anti-fundamental chirals form a fundamental hyper multiplet, however supersymmetry is explicitly broken down to $\mathcal{N}=2$ by mass parameters and superpotential terms which we do not discuss.

It was pointed out in \cite{Aganagic:2013tta} (see also the review \cite{Aganagic:2014kja} and \cite{Aganagic:2014oia,Aganagic:2015cta} for the $ADE$ generalization) that the matrix model above manifestly matches a free boson correlator of \mbox{$q$-Virasoro} screening currents and vertex operators. In fact, by using the results reviewed in the previous subsection we can immediately identify\footnote{\label{qconst}Here we have to use that the $q$-constant $c_\beta(\ul w;q)$ in (\ref{Deltabeta}) simply contributes with a constant multiplicative factor to the integral along the contour chosen in \cite{Aganagic:2013tta}.}
\be\label{Bcorr}
\mathcal{B}^{\rm 3d}_c=\oint_{c}\rd^N \ul{w}\; \bra{\alpha_\infty}\prod_{f=1}^{N_{\rm f}} {\sf H}_{\gamma_f}(z_f)\prod_{j=1}^N \S(w_j)\ket{\alpha_0}~,
\ee
up to proportionality factors, provided that we identify
\be
\begin{array}{|c|c|c|c|c|c|}
\hline
~\textrm{Gauge theory}~&~q~&~m_{\rm a}~&m_f~&~\bar m_f~&~\kappa_1\\
\hhline{|======|}
~\textrm{$q$-Virasoro}~&~q~&~t~&~q^{\frac{1}{2}}t^{\frac{\gamma_f}{2}}z_f^{-1}~&~q^{-\frac{1}{2}}t^{-\frac{\gamma_f}{2}}z_f^{-1} ~&~\sqrt{\beta}(\alpha_0+\sqrt{\beta}N-Q_\beta)\\
\hline
\end{array}~,
\ee
with $\alpha_\infty=\alpha_0+2\sqrt{\beta}N+\sum_f\gamma_f$. We see that the screening currents provide the vector and adjoint integration measure, while each vertex operator provide the 1-loop potential of a pair of fundamental/anti-fundamental chiral multiplets (the letter ${\sf H}$ refers indeed to hyper multiplet, while the half vertex operator $\V$ introduced in (\ref{Vvertex}) couples a single chiral).  

The correlator (\ref{Bcorr}) is a particular projection of the state given in (\ref{qvirgen}): we have simply to shift the time variables $\tau_n$ as in (\ref{qtaushift}) and then set $\tau_n=0$. Therefore the natural question arises whether we can give a gauge theory interpretation to the more general state (\ref{qvirgen}). In order to answer to this question we have to include more observables in the gauge theory. One kind of such observable is given by supersymmetric Wilson loops along the $S^1$ at the tip of the disk.\footnote{\label{thetadefects}Other defects includes boundary degrees of freedom such as walls, see e.g. \cite{Gadde:2013wq}. We will consider such insertion in section \ref{sec:CS} when discussing the inclusion of CS terms.} Such insertions contribute to the integral kernel $\Upsilon^{\rm 3d}(\ul w)$ with 
\be
{\rm Tr}_\mathcal{R}\left(\bs w\right)=s_\mathcal{R}(\ul w)~,
\ee 
where $\mathcal{R}$ is an arbitrary ${\rm U}(N)$ representation, $s_\mathcal{R}(\ul w)$ is the associated Schur polynomial, $\bs w=\prod_j w_j^{\bs h^j}$ is the Wilson line at the localization locus and $\{\bs{h}^j,j=1,\ldots, N\}$ are Cartan generators of the gauge group. By using the Cauchy identity 
\be\label{schurid}
\sum_\mathcal{R} s_\mathcal{R}(\hat{\ul \tau})s_\mathcal{R}(\ul w)=\e^{-\sum_{j,k}\ln(1-\hat\tau_k w_j )}=\e^{\sum_{n>0}\tau_n \sum_j w_j^n }~,\quad \tau_n=\sum_k\frac{\hat\tau^n_k}{n}~,
\ee
we are naturally led to package Wilson loop v.e.v.'s into the generating function
\begin{multline}\label{d2xs1gen}
Z(\ul \tau)=\sum_{\mathcal{R}}s_\mathcal{R}(\hat{\ul \tau})\oint_c\frac{\rd^N \ul w}{2\pi\i \ul w}\prod_{j=1}^N w_j^{\kappa_1}\prod_{k\neq j=1}^N\frac{(w_k w_j^{-1};q)_\infty}{(m_{\rm a} w_k w_j^{-1};q)_\infty}\; s_\mathcal{R}\left(\ul w\right)=\\
=\oint_c\frac{\rd^N \ul w}{2\pi\i \ul w}\;\Delta_\beta(\ul w)\;\e^{\sum_{n>0}\tau_n\sum_j w_j^n+\kappa_1\sum_j\ln w_j}~,
\end{multline}
which matches (modulo the remark in footnote \ref{qconst}) the highest weight state (\ref{qvirgen}) by using the representation (\ref{qviriso}). Our interpretation also implies that the generating function of the theory on $D^2\times S^1$ satisfies the \mbox{$q$-Virasoro} constraints (\ref{qvirconstraints}).

Starting from the next section, we are going to study 3d $\mathcal{N}=2$ gauge theories on compact spaces. We are going to show that the \mbox{$q$-Virasoro} algebra still plays a prominent role, but we have to introduce a new remarkable structure: the modular double. 

%After the redefinition
%\be
%\hat\a_n=\sqrt{\frac{2 n^2}{(q^\frac{n}{2}-q^{-\frac{n}{2}})(t^\frac{n}{2}-t^{-\frac{n}{2}})(p^\frac{n}{2}+p^{-\frac{n}{2}})}}\;\a_n~\quad\Rightarrow\quad [\hat\a_n,\hat \a_m]=2n\delta_{n+m,0}~,
%\ee
%and upon setting $q=\e^{\hbar}$, it can be verified that
%\be
%\T_n=2\delta_{n,0}+\hbar^2 \beta \left({\sf L}_n+\frac{Q_\beta^2}{4}\delta_{n,0}\right)+O(\hbar^4)~,
%\ee
%where the Virasoro generators ${\sf L}_n$ are defined in (\ref{Virgen}) by substituting $\a_n\to \hat \a_n$.

\section{Gauge theory on $S^3_b$}\label{sec:S3}
Supersymmetric gauge theories can be conveniently studied on compact spaces, and 3d $\mathcal{N}=2$ YM-CS theories can be placed in a variety of backgrounds while preserving 2 supercharges of opposite R-charge  \cite{Festuccia:2011ws,Dumitrescu:2012ha,Closset:2012ru,Closset:2013vra,Closset:2014uda}. Expectation values of supersymmetric observables computed through Coulomb branch localization provide interesting examples of matrix models, and our goal is to show that we can use \mbox{$q$-Virasoro}/W algebras techniques to study these theories. We will be focusing on single node ${\rm U}(N)$ theories coupled to 1 adjoint chiral multiplet and possibly (anti-)fundamental chirals, postponing the discussion of more general unitary quiver theories to section \ref{sec:generalizations}. We will not discuss superpotential terms which cannot be seen by the matrix model except for possible restrictions on the parameters of the theory.

In this section we focus on gauge theories on the squashed $S^3_b$ geometry \cite{Kapustin:2009kz,Hama:2011ea,Imamura:2011wg,Alday:2013lba} (see also the review \cite{Hosomichi:2014hja}). The $S^3_b$ can be defined by the usual embedding $S^3\subset \mathbb{R}^4$ endowed with the metric 
\be\label{S3b}
\rd s^2= \omega_1^2(\rd x_1^2+\rd x_2^2)+\omega_2^2(\rd x_3^2+\rd x_4^2)~,\quad b^2=\frac{\omega_2}{\omega_1}~,\quad \omega=\omega_1+\omega_2~,
\ee
where $\omega_{1,2}$ are squashing parameters.\footnote{In this parametrization the squashing parameters are assumed to be real. However, once the gauge theory observables are computed by localization, $\omega_{1,2}$ can be taken to be complex.} For our purposes it is useful to keep in mind that $S^3_b$ can be obtained by gluing two solid tori $D^2\times S^1$ through the $S\in{\rm SL}(2,\mathbb{Z})$ element acting on the boundary torus with modular parameter $\epsilon$, namely
\be
q=\e^{2\pi\i\epsilon}\to \e^{-2\pi\i S\cdot \epsilon}~,\quad \epsilon \to S\cdot\epsilon =-\frac{1}{\epsilon}~.
\ee

\subsection{Generating function}
%The matrix model (\ref{ZNS3}) is quite obiquitous in the study of 3d $\mathcal{N}=2$ supersymmetric gauge theories. In fact, upon setting $\tau_{k,i}=0$, the matrix model (\ref{ZNS3}) corresponds to the $S^3_b$ partition function \cite{Hama:2011ea} of supersymmetric ${\rm U}(N)$ YM theory coupled to 1 adjoint chiral multiplet with real mass $m=-\i \omega(\beta+\Delta /2)$, where $\Delta$ is the Weyl dimension, and FI term.\footnote{To compare with the literature we should use $S_2(\omega/2-\i X|\omega_1,\omega_2)=s_b(X)$, where $\omega_1=\omega_2^{-1}=b$.} More generally, v.e.v.'s of  supersymmetric observables can be computed by inserting suitable functions of $X$ in the matrix model. In this case we have to consider infinitely many coupling constants or time variables. 
Coulomb branch localization implies that the path integral localizes onto trivial field configurations except for a constant profile of the adjoint vector multiplet scalar $\bs X=\sum_j X_j \bs h^j$ in the Cartan subalgebra, which is to be integrated over. If we consider the YM-CS theory coupled to 1 adjoint chiral multiplet of complexified mass $M_{\rm a}$\footnote{The real mass is $M^{\mathbb{R}}_{\rm a}=\i M_{\rm a}-\i\frac{\omega}{2}\Delta$, where $\Delta$ is the Weyl dimension. The latter is absorbed into the complex mass due to holomorphy \cite{Jafferis:2010un}.} the partition function of the theory is given by\footnote{To compare with the literature we have to use $S_2(\omega/2-\i X|\ul \omega)=s_b(X)$, where $\omega_1=\omega_2^{-1}=b$.}
\be\label{ZNS3}
Z=\mathcal{N}_0\int_{\i\mathbb{R}^N}\!\rd^N \ul X\; \Delta_S(\ul X)\;\e^{\sum_j V(X_j)}~,
\ee
\be\label{Deltaomega}
\Delta_S(\ul X)=\prod_{k\neq j}\frac{S_2(X_k-X_j|\ul\omega)}{S_2(M_{\rm a}+X_k-X_j|\ul\omega)}~,
\ee 
where $\Delta_S(\ul X)$ is the integration measure capturing the vector and adjoint contributions and the double Sine function is defined in (\ref{S2true}). The potential $V(X)$ is determined by the classical CS action (including the FI) 
\be
V(X)=
-\frac{\i\pi\kappa_2}{\omega_1\omega_2} X^2 +\frac{2\pi\i\kappa_1}{\omega_1\omega_2} X~,
\ee
where $\kappa_2$ is the CS level and $\kappa_1$ is the FI parameter, while $\mathcal{N}_0$ is an overall normalization constant. Notice that the theory with $\kappa_2=\kappa_1=0$ is a.k.a. the $\mathcal{N}=2^*$ theory, namely the pure $\mathcal{N}=4$ YM theory broken down to $\mathcal{N}=2$ by the adjoint mass. % We will show that the partition function (\ref{ZNS3}) satisfies two commuting copies of \mbox{$q$-Virasoro} constraints. 

A relevant class of observables which can be computed though localization is provided by supersymmetric Wilson loops in arbitrary representations of the gauge group, which can be inserted along the great circles $x_3=x_4=0$ (length $2\pi/\omega_1$) or $x_1=x_2=0$ (length $2\pi/\omega_2$) at the North or South poles of the Hopf base (see for instance \cite{Tanaka:2012nr}). Such insertions amount to evaluate the v.e.v. of 
%\cite{Fiol:2013hna,Dubinkin:2013tda}
\be
{\rm Tr}_{\mathcal{R}_i}\left(\e^{\frac{2\pi\i}{\omega_i}{\bs X}}\right)=s_{\mathcal{R}_i}(\e^{\frac{2\pi\i}{\omega_i} \ul X})~,
\ee 
for an arbitrary ${\rm U}(N)$ representation  $\mathcal{R}_i$. Using the identity (\ref{schurid}) we are naturally led to package Wilson loop v.e.v.'s into the generating function
\begin{multline}\label{ZNS3}
Z(\ul\tau_{1},\ul\tau_{2})=\sum_{\mathcal{R}_1,\mathcal{R}_2} \mathcal{N}_0\int_{\i\mathbb{R}^N}\!\rd^N \!\ul X\; \Delta_S(\ul X)\;\e^{\sum_j V(X_j)} \prod_{i=1,2}s_{\mathcal{R}_i}(\hat{\ul\tau}_{i})s_{\mathcal{R}_i}(\e^{\frac{2\pi\i}{\omega_i}\ul X})=\\
=\int_{\i\mathbb{R}^N}\!\rd^N\! \ul X\; \Delta_S(\ul X)\;\e^{\sum_j V(X_j)}\prod_{i=1,2} \exp\left(\sum_{n> 0}\tau_{n,i} \sum_j \e^{\frac{2\pi\i n}{\omega_i}X_j}+\tau_{0,i}N\kappa_0 \right) ~,
\end{multline}
where we parametrized $\mathcal{N}_0=\exp{N\kappa_0 (\tau_{0,1}+\tau_{0,2})}$. Therefore, the inclusion of Wilson loops modifies the matrix model potential according to 
\be
V(X)\to V(X|\ul \tau_1,\ul \tau_2)=V(X)+\sum_{i=1,2}\left(\sum_{n> 0}\tau_{n,i}  \e^{\frac{2\pi\i n}{\omega_i}X}+\tau_{0,i}\kappa_0 \right)~.
\ee
Notice that the partition function $Z$ is simply $Z(\ul 0,\ul 0)$. However, the generating function is a much more interesting object to study as it contains more information than the bare partition function. Moreover, the contribution of (anti-)fundamental matter can be included as a background for the time variables and CS levels, which is the reason why we will be mostly interested in the theory without (anti-)fundamental matter. In fact, a fundamental chiral multiplet of complexified mass $M_{\rm f}$ can can be coupled to the theory by adding the following 1-loop term to the potential 
\be
V(X)_{\rm fund.}=-\ln S_2(X+M_{\rm f}|\ul\omega)~.
\ee
On the other hand, for generic squashing parameters (i.e. ${\rm Im}\left(\omega_2/\omega_1\right)\neq 0$) the double Sine function has the representation (\ref{S2})
\be
-\ln S_2(X|\ul{\omega})=-\frac{\i\pi}{2\omega_1\omega_2}\left(X^2-X\omega+\frac{\omega^2+\omega_1\omega_2}{6}\right)+\sum_{i=1,2}\sum_{n>0}\frac{\e^{\frac{2\pi\i n}{\omega_i}X}}{n(1-\e^{2\pi\i n\frac{\omega}{\omega_i}})}~.~~~
\ee
It is therefore clear that a fundamental chiral multiplet can be simply coupled to the theory by shifting the time variables and CS levels according to
\begin{align}
\kappa_2&\to \kappa_2+\frac{1}{2}~,\quad \kappa_1\to \kappa_1+\frac{\omega}{4}-\frac{M_{\rm f}}{2}~,\quad \tau_{n,i}\to \tau_{n,i}+\frac{\e^{\frac{2\pi\i n}{\omega_i}M_{\rm f}}}{n(1-\e^{2\pi\i n\frac{\omega}{\omega_i} })}~,\quad n>0~,
\end{align}
and $\ln\mathcal{N}_0\to \ln\mathcal{N}_0-\frac{\i\pi}{2\omega_1\omega_2}(M_{\rm f}^2-M_{\rm f}\omega-\frac{\omega_1^2+\omega_2^2-3\omega^2}{12})$. It is worth noting that the shift of the time variables alone corresponds to gauging a tetrahedron theory \cite{Dimofte:2011ju,Dimofte:2011py} for each weight, which automatically gets rid of the parity anomaly due to half-integer CS units. 

In the next subsection we are going to show that the generating function (\ref{ZNS3}) has a neat \mbox{$q$-Virasoro} interpretation and that it satisfies two commuting copies of \mbox{$q$-Virasoro} constraints.

\subsection{Free boson realization}
%This generating function coincides with the matrix model (\ref{ZNS3}) up to proportionality factors. Viceversa, the existence of two commuting sets of \mbox{$q$-Virasoro} constraints suggests that the generating function can be factorized according to \FN{This is symbolic, the meaning has to be clarified.}
%\be
%W(\{\tau_{1,i},\tau_{2,i}\})=\sum_{\{a\}}\langle \exp\left(\sum_{k> 0}\tau_{1,i} \sum_j \e^{\frac{2\pi\i k}{\omega_1}X_j} \right)\ket{a}\bra{a}\exp\left(\sum_{k> 0}\tau_{2,i} \sum_j \e^{\frac{2\pi\i k}{\omega_2}X_j} \right)\rangle~.
%\ee
To begin with, let us consider the matrix model
%\be\label{ZNS3bis}
%\begin{split}
%Z(\ul\tau_{1},\ul\tau_{2})&=\int_{\i\mathbb{R}^N}\!\rd^N\! \ul X\; \Delta_S(\ul X)\;\e^{\sum_j V(X_j|\ul\tau_{1},\ul\tau_{2})}~,\\
%V(X|\ul\tau_{1},\ul\tau_{2})&=\frac{2\pi\i\kappa_1}{\omega_1\omega_2} X +\sum_{i=1,2}\left(\sum_{n> 0}\tau_{n,i}\e^{\frac{2\pi\i n}{\omega_i}X}+\tau_{0,i}\kappa_0   \right)~,
%\end{split}
%\ee
corresponding to the gauge theory generating function (\ref{ZNS3}) without pure CS action, i.e. $\kappa_2=0$. We discuss the inclusion of CS terms in section \ref{sec:CS}. As reviewed in section \ref{sec:free_field}, the simplest strategy to derive the constraints satisfied by the matrix model is to look for its free boson realization. We  can in fact  give such representation by means of two commuting copies of the very same Heisenberg algebra (\ref{qosc}) (we display non-trivial relations only)
\be\label{qosc2}
\!\!\!\!\!\!\!\!\! [\a_{n,i},\a_{m,i}]=\frac{1}{n}(q_i^\frac{n}{2}-q_i^{-\frac{n}{2}})(t_i^\frac{n}{2}-t_i^{-\frac{n}{2}})(p_i^\frac{n}{2}+p_i^{-\frac{n}{2}})\delta_{n+m,0}~,\quad \!\!\! [\P_i,\Q_i]=2~, \quad\!\!\! n,m\in\mathbb{Z}\backslash\{0\}~,
\ee
where the subindex $i=1,2$ denotes the two copies. The \mbox{$q$-Virasoro} and gauge theory parameters are related by
\be\label{Sgluing}
\begin{array}{|c|l|l|}
\hline
\textrm{~Gauge}~&\multicolumn{2}{c|}{\omega_1~,\omega_2~,M_{\rm a}}\\
\hhline{|=|==|}
\multirow{4}{*}{\textrm{~$q$-Virasoro}}~&\quad\quad\quad\textrm{Copy 1}&\quad\quad\quad\textrm{Copy 2}\\
\cline{2-3}
&~q_1=\e^{2\pi\i\frac{\omega}{\omega_1}}~&~ q_2=\e^{2\pi\i\frac{\omega}{\omega_2}}~\\
&~t_1=\e^{2\pi\i\frac{\beta_1\omega}{\omega_1}}=\e^{\frac{2\pi\i}{\omega_1}M_{\rm a}}~&~ t_2=\e^{2\pi\i\frac{\beta_2\omega}{\omega_2}}=\e^{\frac{2\pi\i}{\omega_2}M_{\rm a}}~\\
&~\beta_1=\beta ~&~ \beta_2=\beta~\\
\hline
\end{array}\quad~.
\ee
By introducing the fundamental weight variables (i.e. the Wilson lines evaluated on the $j^{\rm th}$ fundamental weight)
\be\label{wS3}
(w_j)_1=\e^{\frac{2\pi\i}{\omega_1}X_j}~,\quad 
(w_j)_2=\e^{\frac{2\pi\i}{\omega_2}X_j}~,
\ee
the integration measure (\ref{Deltaomega}) is reproduced by the current 
\be\label{S:S}
\mathcal{S}(X)=(w)_1(w)_2\; \S(w)_1\otimes\S(w)_2~,
\ee
which is essentially the product of two commuting \mbox{$q$-Virasoro} screening currents defined in (\ref{SqVir}).
%\footnote{Here we are using the tensor product symbol to emphasize the commutativity of the two factors.} 
 In fact 
\be\label{SSS:S}
\prod_j \mathcal S(X_j)=\; :
\prod_j \S(w_j)_1\otimes\S(w_j)_2:\Delta_S(\ul X)\; \e^{\frac{2\pi\i\omega\sqrt{\beta}}{\omega_1\omega_2}(\sqrt{\beta}N-Q_\beta)\sum_j X_j}~,
\ee
where we recall the definition $Q_\beta=\sqrt{\beta}-1/\sqrt{\beta}$. Here we used (\ref{Deltabeta}), the representation (\ref{S2}) of the double Sine function and the modular property (\ref{ThetaS}) of the $\Theta$ function with $\epsilon=\omega/\omega_1$, $r=1$. Table (\ref{Sgluing}) summarizes the $g\in {\rm SL}(2,\mathbb{Z})$ gluing involved in our construction
\be\boxed{
\epsilon\to g\cdot \epsilon=\frac{\epsilon}{1-\epsilon}\quad \Rightarrow\quad q_1=\e^{2\pi\i\epsilon}~,\quad q_2=\e^{-2\pi\i g\cdot \epsilon}}~,
\ee
which nicely reflects the geometric decomposition of $S^3_b$ into a pair of solid tori $D^2\times S^1$ each equipped with its own copy of the \mbox{$q$-Virasoro} algebra. Moreover, by parametrizing $X/\omega_1=\chi$ we also have the $g\in {\rm SL}(2,\mathbb{Z})$ action on the coordinates 
\be
\boxed{
\chi\to g\cdot \chi = \frac{\chi}{1-\epsilon}\quad \Rightarrow \quad (w)_1=\e^{2\pi\i\chi}~,\quad (w)_2=\e^{-2\pi\i g\cdot \chi}}~.
\ee
We should notice that the above parametrization is adapted to the lens space description $S^3_b\simeq L(1,1)$ of the next section, but we can recover the more familiar description through the $S\in {\rm SL}(2,\mathbb{Z})$ gluing by identifying $\epsilon\simeq \epsilon+1$.

We next define the  operator 
\be\label{Qcharge}
\mathcal{Z}=\mathcal{J}^N~,\quad \mathcal{J}=\int_{\i\mathbb{R}} \!\rd X\;\mathcal{S}(X)~
\ee
built from the screening charge $\mathcal{J}$. When acting on the charged Fock vacuum state $\ket{\alpha}$ defined by
\be
\ket{\alpha}=\e^{\frac{\alpha}{2}\Q_1}\otimes\e^{\frac{\alpha}{2}\Q_2}\ket{0}~,\quad \a_{n,i}\ket{0}=0~,\quad \P_i\ket{\alpha}=\alpha\ket{\alpha}~,\quad n>0~,
\ee
we get
\begin{multline}
\mathcal{Z}\ket{\alpha}=\int_{\i\mathbb{R}^N}\!\rd^N \ul X\; \Delta_S(\ul X)\;\e^{\frac{2\pi\i \omega\sqrt{\beta}}{\omega_1\omega_2}(\alpha+\sqrt{\beta}N-Q_\beta)\sum_j X_j}\; \times\\
\times\bigotimes_{i=1,2}\exp\left(\sum_{n>0}\frac{\sum_j (w_j^n)_i}{q_i^{n/2}-q_i^{-n/2}}\;\a_{-n,i}\right)\e^{\sqrt{\beta}N\Q_i}\ket{\alpha}~.
\end{multline}
Using the algebra representation  (\ref{qviriso}) we can finally write 
\begin{multline}\label{ZNS3op}
\mathcal{Z}\ket{\alpha}\simeq Z(\ul\tau_{1},\ul\tau_{2})= \int_{\i\mathbb{R}^N}\rd^N\! \ul X\; \Delta_S(\ul X)\;\e^{\frac{2\pi\i \omega\sqrt{\beta}}{\omega_1\omega_2}(\alpha+\sqrt{\beta}N-Q_\beta)\sum_j X_j}\; \times\\
\times\prod_{i=1,2}\exp\left(\sum_{n>0}\tau_{n,i}\sum_j \e^{\frac{2\pi\i n}{\omega_i}X_j}+\tau_{0,i}\left(N+\frac{\alpha}{2\sqrt{\beta}}\right)\right)~,
\end{multline}
matching the gauge theory generating function provided that we identify 
\be\label{kappa:param:S3}
\kappa_1=\omega\sqrt{\beta}(\alpha+\sqrt{\beta}N-Q_\beta)~,\quad \kappa_0=1+\frac{\alpha}{2N\sqrt{\beta}}~.
\ee
In order to show that the matrix model (\ref{ZNS3op}) satisfies two commuting copies of \mbox{$q$-Virasoro} constraints, we have to verify that the \mbox{$q$-Virasoro} generators $\T_{n,i}$ in the free boson representation (\ref{qosc2}) commute with the screening current (\ref{S:S}) up to total differences, namely 
\be\label{TS:S}
[\T_{n,i},\mathcal{S}(X)]=\mathcal{O}(\lambda_i+X)_i-\mathcal{O}(X)_i~
\ee
for some ($n$-dependent) operator $\mathcal{O}(X)_i$ and $\lambda_i\in\mathbb{C}$. Indeed, assuming that $\sqrt{\beta}\P_i$ has integer eigenvalues,\footnote{We can relax this constraint by modifying the zero modes without affecting the algebra.} we can use the $\omega_i$-periodicity in the $i^{\rm th}$ copy  and the relation (\ref{TS}) valid for the two copies
\be\label{TSi:1}
\begin{split}
[\T_{n,1},\S(w)_i]&=(w)_i^{-1}\left({\sf O}(q_i w)_i-{\sf O}(w)_i\right)\delta_{1,i}~,\\
[\T_{n,2},\S(w)_i]&=(w)_i^{-1}\left({\sf O}(q_i w)_i-{\sf O}(w)_i\right)\delta_{2,i},~
\end{split}
\ee
to conclude that (\ref{TS:S}) holds true with
\be\label{TSi}
\mathcal{O}(X)_1=(w)_2\, {\sf O}(w)_1\otimes \S(w)_2~,\quad \mathcal{O}(X)_2=(w)_1\, \S(w)_1\otimes {\sf O}(w)_2~,\quad \lambda_{1,2}=\omega_{2,1}.~~~~~~
\ee
This non-trivial property of the screening current $\mathcal{S}(X)$ and the ${\rm SL}(2,\mathbb{Z})$ pairing justify the name ``modular double" that we are using for our construction. Finally, from the interpretation (\ref{ZNS3op}) and the highest weight condition
\be
\T_{n,i}\ket{\alpha}=0~,\quad n>0~,
\ee 
we easily get \mbox{$q$-Virasoro} constraints through the representation (\ref{qviriso}) of (\ref{qvirconstraints})
\be
T(z|\ul\tau_i)_i \; Z(\ul \tau_1,\ul \tau_2)={\rm Pol}(z)_i\quad \Rightarrow\quad T_{n,i}(\ul \tau_i)\; Z(\ul \tau_1,\ul \tau_2)=0~,\quad n>0~,\nn
\ee
for a certain ($\tau_i$-dependent) polynomial ${\rm Pol}(z)_i$ and where $T_{n,i}(\ul \tau_i)$ are the differential operators given in (\ref{TnBell}).

%The discussion of this section can be compared for instance with the results of \cite{Bullimore:2014awa}, where the $S^3_b$ partition function of the $T[{\rm U}(N)]$ theory is identified with the joint eigenfunction of two commuting sets of difference operators mapped to tRS Hamiltonians/'t Hooft loop operators in 4d and whose eigenvalues are flavor Wilson loops.

\subsection{Round $S^3$: W$_{1,t}$ and Virasoro limits}\label{sec:limits}
The $q\textrm{-Virasoro}=\textrm{W}_{q,t}(A_1)$ algebra admits several interesting limits in which it reduces to other known algebras, the most famous one being the conformal limit discussed around (\ref{virlimit}). Other interesting limits are: the Hall-Littlewood limit $q\to 0$ with $t$ fixed \cite{Awata:1996fq} and recently discussed in \cite{Ohkubo:2015roe} in the context of the 5d AGT correspondence; the root of unity limit \cite{Bouwknegt:1997} recently discussed in the context of the 4d AGT correspondence in \cite{Itoyama:2013mca,Itoyama:2014pca}; the special values $\beta=1,3/2,2$ in which case connections with Kac-Moody, topological and $\textrm{W}_{1+\infty}$ algebras respectively were discussed in \cite{Awata:1996fq}; the Frenkel-Reshetikhin limit $t\to 1$ with $q$ fixed (classical \mbox{$q$-Virasoro} algebra) or $q\to 1$ with $t$ fixed \cite{1997q.alg.....8006F}, in which case the algebra becomes commutative but inherits a natural Poisson algebra structure isomorphic to the Poisson algebra obtained from the difference Drinfeld-Sokolov reduction of $\widehat{\rm SL}_2$ \cite{Frenkel1998,Semenov-Tian-Shansky1998}.

It would be very interesting to understand all these limits from the viewpoint of the \mbox{$q$-Virasoro} modular double and 3d gauge theories on compact spaces, but the general discussion is beyond the aim of this work. Also, we should observe that taking the limits at the algebra level might be very subtle: in fact, it may happen that a particular (naive) limit on one \mbox{$q$-Virasoro} factor is ill-defined on the other. In the following we will simply ignore these subtleties and study instead a couple of particular limits where the compact space generating function is perfectly defined and allows us to explicitly find the constraints it satisfies by standard matrix model techniques. For concreteness, we will focus on the $S^3_b$ geometry discussed in this section, analyzing  the matrix model (\ref{ZNS3op}) in special limits of the deformation parameters. First of all, we consider the round $S^3$ limit corresponding to $\omega_1\to\omega_2\to 1$ from a complex direction. In terms of the parameters of the \mbox{$q$-Virasoro} algebra this limit corresponds to $q_1^{1/2}\to q_2^{1/2}\to 1$. However, in this limit the value of $t_1^{1/2}\to t_2^{1/2}\to\e^{2\pi\i\beta}$ is still a free parameter and each copy of the \mbox{$q$-Virasoro} algebra should reduce to the W$_{1,t}(A_1)$ algebra mentioned above. In order to be able to study the matrix model exactly, we can take a further limit on $\beta$ such that $t_1\to t_2\to \pm 1$, in which case we expect to find a relation with the Virasoro algebra (a different one w.r.t. (\ref{virlimit}) though).

\subsubsection*{$\beta\in\frac{\mathbb{Z}}{2}$ or $t=1$: Virasoro limit}

The first simple example is when $\beta\in\frac{\mathbb{Z}}{2}$, in which case $t_{1,2}\to t=\e^{4\pi i \beta}\to 1$.
%, and following the discussion at the end of section \ref{sec:free_field} we expect that the \mbox{$q$-Virasoro} constraints reduce to the usual Virasoro constraints. 
In this limit the measure (\ref{Deltaomega}) of the matrix model (\ref{ZNS3op}) reduces to the following expression 
\be\label{Vir:round}
\Delta_S(\ul X)= \left((-1)^{\beta(2\beta-1)}2^{2\beta}\right)^{N(N-1)}\prod_{k\neq j} \sin^{2\beta}\left(\pi(X_k-X_j)\right)~,
\ee
which can be derived from the reflection property (\ref{reflection}). Therefore the matrix model becomes 
\begin{multline}
Z(\ul \tau)=\int_{\i\mathbb{R}^N} \!\rd^N \ul X \prod_{k\neq j}\sin^{2\beta}\left(\pi(X_k-X_j)\right)\times\\
\times\e^{2\pi \i\kappa_1\sum_j X_j}
\exp\left(\sqrt{2\beta}\sum_{n>0}\sum_j \tau_n \e^{2\pi \i n X_j}+\sqrt{2\beta}\tau_0 N \kappa_0\right)~,
\end{multline}
where we set $\sqrt{2\beta}\tau_n=\tau_{n,1}+\tau_{n,2}$ and neglected an overall prefactor  which is not relevant 
for our further discussions. Introducing the exponentiated variables \mbox{$x_j=\e^{2\pi \i X_j}$}, we can rewrite the matrix model as
\be\label{partition:q=1:b=1/2}
Z(\ul \tau)= \int_{0}^{\infty} \!\rd^N\! \ul x \prod_j x_j^{(1+2\sqrt{\beta}\alpha)}
\prod_{k\neq j}(x_k-x_j)^{2\beta}\e^{\sqrt{2\beta}(\sum_{n>0} \tau_n\sum_j x_j^n+\tau_0 N\kappa_0)}~,
\ee
up to unimportant proportionality factors, and where we used $\kappa_1=2\sqrt{\beta}(\sqrt{\beta}N-Q_\beta+\alpha)$. Now we can derive the constraints for this matrix integral using standard techniques (for details we refer to \cite{Mironov:1990im,Nedelin:2015mio,Dijkgraaf:1990rs}). We shift the integration variables 
$x_j\to x_j+\varepsilon_n x_j^{n+1}$, $n\in\mathbb{Z}_{>0}$, and collect all the variations under the integral, leading to the following 
Ward identities 
\be
\langle \sum_{k\geqslant 1}\sqrt{2\beta}\tau_k k\sum_j x_j^{n+k}+2\beta\sum\limits_{k=1}^{n-1}\sum_{\ell,j}x_\ell^k x_j^{n-k}+\left(4\beta N+2\sqrt{\beta}\alpha +1+(n+1)(1-2\beta)\right)\sum_j x_j^n\rangle=0~,
\ee
where $\langle~\rangle$ denotes the matrix model average. Such identities are equivalent to the differential constraints $L_n(\ul \tau) Z(\ul \tau)=0$ where the $L_n(\ul \tau)$ operators are 
\be
L_n(\ul \tau)=\sum_{k\geqslant 1}\tau_k k\frac{\partial}{\partial \tau_{n+k}}+\sum_{k=1}^{n-1}
\frac{\partial^2}{\partial \tau_k\partial \tau_{n-k}}+\left(2\sqrt{2\beta} N+\sqrt{2}\alpha +\frac{1}{\sqrt{2\beta}}-(n+1)Q_{2\beta}\right)\frac{\partial}{\partial \tau_n}~.
\ee
We can also use
\be
2\frac{\partial}{\partial \tau_0}Z(\ul \tau)=2\sqrt{2\beta}N\kappa_0Z(\ul \tau)~,\quad \kappa_0=1+\frac{\alpha}{2N\sqrt{\beta}}~,
\ee
to rewrite
\be
L_n(\ul \tau)=\sum_{k\geqslant 1}\tau_k k\frac{\partial}{\partial \tau_{n+k}}+\sum_{k=1}^{n-1}
\frac{\partial^2}{\partial \tau_k\partial \tau_{n-k}}+\left(2\frac{\partial}{\partial \tau_0}+\frac{1}{\sqrt{2\beta}}-(n+1)Q_{2\beta}\right)\frac{\partial}{\partial \tau_n}~.
\ee
These differential operators represent the Virasoro operators (\ref{Lop}) with zero mode $\P\simeq 2\frac{\partial}{\partial \tau_0}+\frac{1}{\sqrt{2\beta}}$
and central charge $c=1-6Q_{2\beta}^2$.

\subsubsection*{$\beta\in\frac{1}{4}+\frac{\mathbb{Z}}{2}$ or $t=-1$: another Virasoro limit}

Another situation in which we can explicitly derive the constraints satisfied by the matrix model is  when 
$\beta\in\frac{1}{4}+\frac{\mathbb{Z}}{2}$. This case corresponds to $t_{1,2}\to t=\e^{4\pi\i\beta}\to -1$.
In order to find the limit of the measure (\ref{Deltaomega}) of the matrix model (\ref{ZNS3op}) we 
use 
\be
\frac{S_2(X+\frac{1}{2}|1,1)}{S_2(X+n+\frac{1}{2}|1,1)}=(-1)^{\frac{n(n -1)}{2}}2^{n}\cos^{n}(\pi X)~,\quad n=2\beta-\frac{1}{2}\in\mathbb{Z}~,
\ee
and the reflection property (\ref{reflection}), leading the following matrix model 
\begin{multline}
Z(\ul \tau)= \int_{\i\mathbb{R}^N} \!\rd^N\! \ul X\prod_{k<j}\sin^2(\pi(X_k-X_j))\cos^{(4\beta-2)}(\pi(X_k-X_j))\times\\
\times \e^{2\pi\i\kappa_1\sum_j X_j}\exp\left(\sqrt{2\beta}\sum_{n>0}\sum_j \tau_n e^{2\pi i n X_j}+\sqrt{2\beta}\tau_0 N \kappa_0\right)~,
\end{multline}
up to proportionality factors, and where we set $\sqrt{2\beta}\tau_n=\tau_{n,1}+\tau_{n,2}$. Rewriting the matrix model in terms of the $x_j=\e^{2\pi\i X_j}$ variables we obtain 
\be
Z(\ul \tau)= \int_0^\infty \!\!\!\rd^N\! \ul x \prod_j x_j^{(1+2\alpha\sqrt{\beta})}\prod_{k<j}(x_k-x_j)^2(x_k+x_j)^{2(2\beta-1)}\e^{\sqrt{2\beta}(
\sum_{n>0} \tau_n\sum_j x_j^n+\tau_0 N\kappa_0)}~,~~~~~
\ee
up to proportionality factors, and where we used $\kappa_1=2\sqrt{\beta}(\sqrt{\beta}N-Q_\beta+\alpha)$. Performing the shift $x_j\to x_j+\varepsilon_n x_j^{n+1}$, $n\in2\mathbb{Z}_{>0}$, and collecting all the variations under the integral we arrive at the Ward identities 
\be
\langle\sqrt{2\beta} \sum_{k\geqslant 1}\tau_k k\sum_j x_j^{n+k}+\sum\limits_{k=0}^n\left(1+(-1)^k(2\beta -1)\right)
\sum_{\ell,j}x_\ell^k x_j^{n-k}
+\left(2\sqrt{\beta}\alpha +2-2\beta\right)\sum_j x_j^n\rangle=0~.
\ee
The reason why these Ward identities can be derived for even $n$ only is that the variation of $\prod_{k\neq j}(x_k+x_j)^{2\beta -1}$ contains 
\be
\sum_{k\neq j}\frac{x_k^{n+1}+x_j^{n+1}}{x_k+x_j}=\sum_{k=0}^{n}\sum_{\ell\neq j}(-1)^k x_\ell^k x_j^{n-k}\,,\quad n\in 2\mathbb{Z}~,
\ee
while for odd $n$ a similar simplification to the r.h.s. does not take place. The derived Ward identities can be generated by the action of the following differential operators on the matrix model
\be
L_n(\ul\tau)=\sum_{k\geqslant 1}\tau_k k\frac{\partial}{\partial \tau_{n+k}}+\frac{1}{2\beta}\sum_{k=1}^{n-1}
\left(1+(-1)^k(2\beta -1)\right)\frac{\partial^2}{\partial \tau_k\partial \tau_{n-k}}
+\left(2\frac{\partial}{\partial\tau_0} +\sqrt{\frac{2}{\beta}}-2\sqrt{\frac{\beta}{2}}\right)\frac{\partial}{\partial \tau_n}~.
\ee
In order to interpret these operators, we can make the following identification between the differential and free boson operators satisfying 
the usual Heisenberg algebra $[\a_n,\a_m]=2n\delta_{n+m,0}$
\be
\begin{split}
\a_n&\simeq 2\sqrt{\frac{1+(-1)^n(2\beta-1)}{2\beta}}\frac{\partial}{\partial \tau_n}~,\quad \a_{-n}\simeq n \sqrt{\frac{2\beta}{1+(-1)^n(2\beta-1)}}\tau_n~,\quad n>0~,\\
\P&\simeq 2\frac{\partial}{\partial \tau_0}+\frac{2}{\sqrt{2\beta}}-2\sqrt{\frac{\beta}{2}}~.
\end{split}
\ee
Under this identification our operators become 
\be
L_n(\ul\tau)\simeq {\sf L}_n=\frac{1}{4}\sum_{k\neq 0,n } :\a_{n-k}\a_k:+\frac{1}{2}\a_n\P~,
\ee
satisfying the Virasoro algebra with the central charge $c=1$ when extended to $n\in\mathbb{Z}$.

\section{Other compact backgrounds}\label{sec:other}
In this section we extend our results to other gauge theory backgrounds. Since the analysis will be quite analogous to the previous one for the $S^3_b$ background, but with the important difference given by non-trivial fundamental groups, we will be more concise in the presentation.

\subsection{$L(r,1)$}\label{sec:lens}
The focus of this subsection is on gauge theories on the squashed lens space $L(r,1)=S^3_b/\mathbb{Z}_r$ \cite{Gang:2009wy,Benini:2011nc,Alday:2012au,Imamura:2012rq,Imamura:2013qxa}. The lens space can be defined as the $S^3_b$ with metric (\ref{S3b}) and the additional $\mathbb{Z}_r$ quotient by the action
\be
x_{1}+\i x_2\to \e^{\frac{2\pi\i}{r}}(x_1+\i x_2)~,\quad  x_{3}+\i x_4\to \e^{-\frac{2\pi\i}{r}}(x_3+\i x_4)~.
\ee
It is also useful to keep in mind that $L(r,1)$ can be obtained by gluing two solid tori $D^2\times S^1$ through the $g_r\in{\rm SL}(2,\mathbb{Z})$ element acting on the boundary torus with modular parameter $\epsilon$, namely
\be
q=\e^{2\pi\i\epsilon}\to \e^{-2\pi\i g_r\cdot \epsilon}~,\quad \epsilon \to g_r\cdot\epsilon =\frac{\epsilon}{1-r\epsilon}~.
\ee

\subsection*{Generating function}
Coulomb branch localization implies that the path integral localizes onto flat connections and a constant profile for the adjoint vector multiplet scalar $\bs X$ in the Cartan, which is to be integrated over. Flat connections are classified by $\pi_1(L(r,1))\simeq\mathbb{Z}_r$, and hence labeled by integers $\ul \ell\in\mathbb{Z}_r^N$ which are to be summed over (we consider unordered sequences). If we consider the YM-CS theory coupled to 1 adjoint chiral multiplet of complexified mass $M_{\rm a}$ 
the partition function of the theory is given by
\be
Z=\mathcal{N}_0\sum_{\ul \ell\in\mathbb{Z}_r^N}\int_{\i\mathbb{R}^N}\!\rd^N \!\ul X\; \Delta_r(\ul X,\ul \ell)\;\e^{\sum_j V(X_j,\ell_j)}~,
\ee
where
\be\label{Delta:r}
\Delta_{r}(\ul X,\ul \ell)=\prod_{k\neq j}\frac{S_{2,-(\ell_k-\ell_j)}(X_k-X_j|\ul\omega)}{S_{2,-(\ell_k-\ell_j)}(M_{\rm a}+X_k-X_j|\ul\omega)}~,
\ee
\be
V(X,\ell)=-\frac{\i\pi\kappa_2}{r\omega_1\omega_2}X^2-\frac{\i\pi\kappa_2}{r}\ell^2+\frac{2\pi\i\kappa_1}{r\omega_1\omega_2}X~.
\ee
Here $\kappa_2$ is the CS level, $\kappa_1$ the FI while the generalized double Sine function is defined in (\ref{genS2true}). Supersymmetric Wilson loops can be inserted along the non-contractible cycles at the North and South poles of the Hopf base. Such insertions amount to evaluate the v.e.v. of 
%\cite{Fiol:2013hna,Dubinkin:2013tda}
\be
{\rm Tr}_{\mathcal{R}_i}\left(\e^{\frac{2\pi\i}{r\omega_i}  \bs X} \e^{s_i\frac{2\pi\i}{r} \bs \ell}\right)~,\quad s_{1,2}=\pm 1~,\quad 
\ee 
where $\e^{\pm\frac{2\pi\i}{r}\bs\ell }$ is the holonomy of the gauge connection along the non-contractible cycle. The generating function thus reads as
\be\label{ZNlens}
\begin{split}
Z(\ul \tau_1,\ul \tau_2)&=\sum_{\ul \ell\in\mathbb{Z}_r^N}\int_{\i\mathbb{R}^N}\!\rd^N \!\ul X\; \Delta_r(\ul X,\ul \ell)\;\e^{\sum_j V(X_j,\ell_j|\ul \tau_1,\ul \tau_2)}~,\\
V(X,\ell|\ul \tau_1,\ul \tau_2)&=V(X,\ell)+\sum_{i=1,2}\left(\sum_{n>0}\tau_{n,i}\e^{\frac{2\pi\i n}{r\omega_i}(X+s_i\omega_i\ell)}+\tau_{0,i}\kappa_0\right)~,
\end{split}
\ee
where we parametrized $\mathcal{N}_0=\e^{N\kappa_0(\tau_{0,1}+\tau_{0,2})}$. A fundamental chiral multiplet of complexified mass $M_{\rm f}$ together with $1/2$ CS units can be coupled to the theory by shifting the time variables according to 
\be
\tau_{n,i}\to \tau_{n,i}+\frac{\e^{\frac{2\pi\i n}{r\omega_i}M_{\rm f}}}{n(1-\e^{2\pi\i n \frac{\omega }{r\omega_i}})}~.
\ee

\subsection{$S^2 \times S^1$ (index)}\label{sec:S2xS1id}
In this subsection we consider gauge theories on the $S^2\times S^1$ background associated to the superconformal index \cite{Imamura:2011su,Kapustin:2011jm,Hwang:2015wna}. The $S^1$ period can be parametrized by $q=\e^{\hbar\epsilon}$. It is also useful to keep in mind that $S^2\times S^1$ can be obtained by gluing two solid tori $D^2\times S^1$ through the $id\in{\rm SL}(2,\mathbb{Z})$ element acting on the boundary torus with modular parameter $\epsilon$
\be
q=\e^{\hbar\epsilon}\to \e^{-\hbar id\cdot\epsilon}~,\quad \epsilon \to id\cdot\epsilon =\epsilon~.
\ee

\subsection*{Generating function}
Coulomb branch localization implies that the path integral localizes onto monopole configurations on $S^2$ labeled by quantized fluxes $\bs \ell$ to be summed over (we consider unordered sequences), and constant gauge holonomy $\bs x$ around $S^1$ to be integrated over the maximal torus. The adjoint vector multiplet scalar is also constant and proportional to the flux. If we consider the YM-CS theory coupled to 1 adjoint chiral multiplet with global fugacity $m_{\rm a}$ 
the partition function of the theory is 
\be
Z=\mathcal{N}_0\sum_{\ul \ell\in \mathbb{Z}^N}\oint_{\mathbb{T}^N}\frac{\rd^N \!\ul x}{2\pi\i \ul x}\; \Delta_{id}(\ul x,\ul \ell)\;\e^{\sum_j V(x_j,\ell_j)}~,
\ee
where
\begin{align}\label{Delta:id}
\Delta_{id}(\ul x,\ul \ell)&=\prod_{k< j}m_{\rm a}^{-(\ell_k-\ell_j)}(1-x_k x_j^{-1}q^{\frac{\ell_k-\ell_j}{2}})(1-x_j x_k^{-1}q^{\frac{\ell_k-\ell_j}{2}})\times\nn\\
&\quad\quad\quad\quad\quad\quad\quad\times\frac{(m_{\rm a}^{-1}x_j x_k^{-1}q^{1+\frac{\ell_k-\ell_j}{2}};q)_\infty(m_{\rm a}^{-1}x_k x_j^{-1}q^{1+\frac{\ell_k-\ell_j}{2}};q)_\infty}{(m_{\rm a} x_k x_j^{-1}q^{\frac{\ell_k-\ell_j}{2}};q)_\infty(m_{\rm a} x_j x_k^{-1}q^{\frac{\ell_k-\ell_j}{2}};q)_\infty}~,\\
%&=\frac{(t;q)_\infty^N}{(q t^{-1};q)_\infty^N}\prod_{i\neq j}t^{-\frac{|\ell_i-\ell_j|}{2}}\frac{(1-x_i x_j^{-1}q^{\frac{|\ell_i-\ell_j|}{2}})(t^{-1}x_j x_i^{-1}q^{1+\frac{|\ell_i-\ell_j|}{2}};q)_\infty}{(t x_i x_j^{-1}q^{\frac{\ell_i-\ell_j}{2}};q)_\infty(t x_j x_i^{-1}q^{\frac{|\ell_i-\ell_j|}{2}};q)_\infty}~,\\
V(x,\ell)&=\kappa_2\ell\ln x+\kappa_1\ln x+\eta_1\ell\ln q~.
\end{align}
Here $\kappa_2$ is the CS level, $\kappa_1$ is the FI and we turned on also the holonomy $q^{\eta_1}$ for the topological ${\rm U}(1)$. Supersymmetric Wilson loops can be supported at the poles of $S^2$, and their evaluation amounts to computing the average of 
\be
{\rm Tr}_{\mathcal{R}_i}\left( {\bs x}^{s_i} q^{-\frac{\bs \ell}{2}}\right)~,\quad s_{1,2}=\pm 1~.
\ee 
The generating function of Wilson loop v.e.v.'s. is therefore
\be\label{ZNS2xS1}
\begin{split}
Z(\ul\tau_1,\ul\tau_2)&=\sum_{\ul \ell\in \mathbb{Z}^N}\oint_{\mathbb{T}^N}\frac{\rd^N\!\ul  x}{2\pi\i \ul x}\; \Delta_{id}(\ul x,\ul \ell)\;\e^{\sum_j V(x_j,\ell_j|\ul\tau_1,\ul\tau_2)}~,\\
V(x,\ell|\ul\tau_1,\ul\tau_2)&=V(x,\ell)+\sum_{n\neq 0}\tau_n x^n q^{-\frac{\ell}{2}|n|}+\kappa_0(\tau_{0,1}+\tau_{0,2})~,
\end{split}
\ee
where we set $\tau_{n,1}=\tau_{n}$, $\tau_{n,2}=\tau_{-n}$ for $n> 0$ and parametrized $\mathcal{N}_0=\e^{N\kappa_0(\tau_{0,1}+\tau_{0,2}) }$. A fundamental chiral multiplet of global fugacity $m_{\rm f}$ together with $1/2$ CS units can be coupled to the theory by shifting the time variables
\be
\tau_n\to \tau_n+\frac{m_{\rm f}^n}{|n|(1-q^n)}~,\quad n\neq 0~.
\ee

\subsection{$S^2\times S^1$ (twisted index)}\label{sec:S2xS1A}
In this subsection we consider gauge theories on the A-twisted $S^2\times S^1$ background leading to the twisted index \cite{Benini:2015noa}. The metric is 
\be
\rd s^2=\rd \theta^2+\sin^2\theta(\rd\phi-2\pi\epsilon\rd y)^2+\rd y^2~,
\ee
where $q=\e^{2\pi\i\epsilon}$ can be interpreted as the angular momentum fugacity. This background is characterized by a flux for the R-symmetry connection. It is also useful to keep in mind that $S^2\times S^1$ can be obtained by gluing two solid tori $D^2\times S^1$ through the $id\in{\rm SL}(2,\mathbb{Z})$ element acting on the boundary torus with modular parameter $\epsilon$
\be
q=\e^{2\pi\i\epsilon}\to \e^{-2\pi\i id \cdot\epsilon}~,\quad \epsilon \to id\cdot\epsilon =\epsilon~.
\ee

\subsection*{Generating function}
Coulomb branch localization implies that the path integral localizes onto monopole configurations on $S^2$ labeled by quantized fluxes $\bs \ell$ to be summed over (we consider unordered sequences), and complexified constant gauge holonomy $\bs x$ around $S^1$ to be integrated over the maximal torus. If we consider the YM-CS theory coupled to 1 adjoint chiral multiplet with fugacity $v$ and R-charge $R$ the partition function of the theory is given by (we refer to \cite{Benini:2015noa} for the J.K. contour)
\be
Z=\mathcal{N}_0\sum_{\ul \ell\in \mathbb{Z}^N}\oint_{\rm J.K.}\frac{\rd^N \!\ul x}{2\pi\i\ul  x}\; \Delta_{A}(\ul x,\ul \ell)\;\e^{\sum_j V(x_j,\ell_j)}~,
\ee
where 
\begin{align}\label{Delta:A}
\Delta_{A}(\ul x,\ul \ell)&=\prod_{k<j}(-1)^{\ell_k-\ell_j-R} q^{-\frac{\ell_k-\ell_j}{2}}(1-x_k x_j^{-1}q^{\frac{\ell_k-\ell_j}{2}})(1-x_j x_k^{-1}q^{\frac{\ell_k-\ell_j}{2}})\times\nn\\
&\quad\quad\times\frac{v^{1-R}(x_k x_j^{-1})^{\ell_k-\ell_j}}{(v x_k x_j^{-1} q^{\frac{1-(1+\ell_k-\ell_j-R)}{2}};q)_{1+\ell_k-\ell_j-R}(v x_j x_k^{-1} q^{\frac{1-(1+\ell_j-\ell_k-R)}{2}};q)_{1+\ell_j-\ell_k-R}}~,\\
%&=\prod_{i<j}(-1)^{\ell_i-\ell_j-R}\prod_{i\neq j}q^{-\frac{|\ell_i-\ell_j|}{2}}\frac{(1-x_i x_j^{-1} q^{\frac{|\ell_i-\ell_j|}{2}})(v x_i x_j^{-1})^{\frac{1+\ell_i-\ell_j-R}{2}}}{(v x_i x_j^{-1}q^{\frac{1-(1+\ell_i-\ell_j-R)}{2}};q)_{1+\ell_i-\ell_j-R}}~,\\
V(x,\ell)&=\kappa_2\ell\ln x+\kappa_1\ln x+\eta_1\ell\ln q~.
\end{align}
Here $\kappa_2$ is the CS level, $\kappa_1$ the FI and we turned on also the holonomy $q^{\eta_1}$ for the topological ${\rm U}(1)$, while the finite $q$-Pochhammer symbol is defined in (\ref{nqPoch}). Supersymmetric Wilson loops wrapping the integral curve of $\partial_y+2\pi\epsilon\partial_\phi$ can be supported at the poles of $S^2$. The evaluation of a Wilson loop amounts to compute the average of
\be
{\rm Tr}_\mathcal{R}\left( \bs x q^{s_i\frac{\bs \ell}{2}}\right)~,\quad s_{1,2}=\pm 1~.
\ee 
The generating function then reads as
\be\label{ZN:S2xS1A}
\begin{split}
Z(\ul \tau_1,\ul \tau_2)&=\sum_{\ul\ell\in \mathbb{Z}^N}\oint_{\rm J.K.}\frac{\rd^N\ul  x}{2\pi\i \ul x}\; \Delta_{A}(\ul x,\ul \ell)\;\e^{\sum_j V(x_j,\ell_j|\ul \tau_1,\ul \tau_2)}~,\\
V(x,\ell|\ul \tau_1,\ul \tau_2)&=V(x,\ell)+\sum_{n\neq 0} \tau_n x^{|n|} q^{-\frac{\ell}{2}n}+(\tau_{0,1}+\tau_{0,2})\kappa_0 ~,
\end{split}
\ee
where we also set $\tau_{n,1}=\tau_{n}$, $\tau_{n,2}=\tau_{-n}$ for $n>0$ and parametrized $\mathcal{N}_0=\e^{N\kappa_0(\tau_{0,1}+\tau_{0,2})}$. A fundamental chiral multiplet with global fugacity $m_{\rm f}$ together with $1/2$ CS units can be coupled to the theory by shifting the time variables\be
\tau_n\to \tau_n+\frac{m_{\rm f}^{|n|}}{|n|(1-q^n)}~, \quad n\neq 0~.
\ee

\subsection*{Free boson realization}
The matrix models arising from the different backgrounds find a unified description in terms of the Virasoro modular double. Let us start by identifying the geometric parameters, fundamental weight variable $w$ associated to the supersymmetric Wilson loops, the adjoint mass, fugacity or R-charge as
\be\label{table:virgauge}
\begin{array}{|c|c|c|c|}
\hline
q\textrm{-Virasoro}~&~L(r,1)~&~S^2\times S^1\textrm{ (index) }~&~S^2\times S^1\textrm{ (twisted index) }\\
\hline
q_1~&\e^{2\pi\i\frac{\omega}{r\omega_1}}~&~\e^{\hbar\epsilon}~&~\e^{2\pi\i\epsilon}\\
\hline
q_2~&\e^{2\pi\i\frac{\omega}{r\omega_2}}~&~\e^{-\hbar\epsilon}~&~\e^{-2\pi\i\epsilon}\\
\hline
~(w)_1~&~\e^{\frac{2\pi\i}{r}\ell}\e^{\frac{2\pi\i}{r\omega_1}X}~&~ \e^{-\hbar\epsilon\frac{\ell}{2}}x~&~ \e^{-2\pi\i\epsilon\frac{\ell}{2}}x\\
\hline
~(w)_2~&~\e^{-\frac{2\pi\i}{r}\ell}\e^{\frac{2\pi\i}{r\omega_2}X} ~&~ \e^{-\hbar\epsilon\frac{\ell}{2}}x^{-1} ~&~ \e^{2\pi\i\epsilon\frac{\ell}{2}}x\\
\hline
~t_1=q_1^{\beta_1}~&~\e^{\frac{2\pi\i}{r\omega_1}M_{\rm a}}~&~ m_{\rm a}~&~  \e^{2\pi\i\epsilon\frac{R}{2}}v\\
\hline
~t_2=q_2^{\beta_2}~&~\e^{\frac{2\pi\i}{r\omega_2}M_{\rm a}}~&~ m_{\rm a}^{-1} ~&~ \e^{-2\pi\i\epsilon\frac{R}{2}}v\\
\hline
~(\beta_1,\beta_2)~&~(\beta,\beta)~&~(\beta,\beta)~&~(\beta,R-\beta)\\
\hline
\end{array}~~~.
\ee
This table summarizes the $g\in {\rm SL}(2,\mathbb{Z})$ gluings involved in our construction. If we denote $\omega/r\omega_1=\epsilon$, $X/r\omega_1=\chi$ for $L(r,1)$ or $x=\e^{2\pi\i\chi}$ for $S^2\times S^1$, then the two copies are related by
\be
\begin{array}{|c|c|c|c|}
\hline
q\textrm{-Virasoro}~&~L(r,1)~&~S^2\times S^1\textrm{ (index) }~&~S^2\times S^1\textrm{ (twisted index) }\\
\hline
~q_1~&\e^{2\pi\i\epsilon}~&~\e^{\hbar\epsilon}~&~\e^{2\pi\i\epsilon}~\\
\hline
~q_2~&\e^{-2\pi\i g\cdot\epsilon}~&~\e^{-\hbar g\cdot \epsilon}~&~\e^{-2\pi\i g\cdot\epsilon}~\\
\hline
~(w)_1~&~\e^{\frac{2\pi\i}{r}\ell}\e^{2\pi\i\chi}~&~\e^{-\hbar\epsilon\frac{\ell}{2}}\e^{2\pi\i\chi}~&\e^{-2\pi\i\epsilon\frac{\ell}{2}}\e^{2\pi\i\chi}\\
\hline
~(w)_2~&~\e^{-\frac{2\pi\i}{r}g\cdot\ell}\e^{-2\pi\i g\cdot\chi}~&~\e^{-\hbar g\cdot\epsilon\; g\cdot\frac{\ell}{2}}\e^{-2\pi\i g\cdot\chi}~&\e^{-2\pi\i g\cdot\epsilon\; g\cdot\frac{\ell}{2}}\e^{-2\pi\i g\cdot\chi}\\
\hline
~g\cdot\epsilon~&~\frac{\epsilon}{1-r\epsilon} ~&~ \epsilon ~&~ \epsilon\\
\hline
~g\cdot\chi~&~\frac{\chi}{1-r\epsilon} ~&~ \chi ~&~ -\chi\\
\hline
~g\cdot \ell~&~ \ell~&~ \ell ~&~  -\ell \\
\hline
\end{array}~~~.
\ee

The matrix models are reproduced by the \mbox{$q$-Virasoro} modular double screening current
\be
\mathcal{S}(\chi)=\sum_{\ell\in\mathbb{F}}(w)_1 (w)_2\; \S(w)_1\otimes \S(w)_2~,
\ee
where we recall that $\mathbb{F}=(\mathbb{Z}_r,\mathbb{Z})$ for $L(r,1)$ and $S^2\times S^1$ respectively. In fact
\be\label{SSS:r}
\prod_{j}\mathcal{S}(\chi_j)= \sum_{\ul \ell\in\mathbb{F}^N}:\prod_{j}\S(w_j)_1\otimes \S(w_j)_2: \Delta(\ul\chi,\ul \ell)\; \prod_j \Delta_0(\chi_j,\ell_j) ~,
\ee
where the measure $\Delta(\ul \chi,\ul \ell)$ is the one appearing in (\ref{Delta:r}), (\ref{Delta:id}), (\ref{Delta:A}) respectively, and 
\be
\Delta_0(\chi_j,\ell_j)=\left\{\begin{array}{lll}
\e^{\frac{2\pi\i\omega\sqrt{\beta}}{r\omega_1\omega_2}(\sqrt{\beta}N-Q_\beta)X_j}&:&L(r,1)\\
q^{-\sqrt{\beta}\ell_j(\sqrt{\beta}N-Q_\beta)}&:&S^2\times S^1\textrm{\small (index) }\\
x_j^{R(N-1)+2}q^{-\ell_j(\beta-\frac{R}{2})(N-1)}&:&S^2\times S^1\textrm{\small (twisted index) }
\end{array}\right.~.~~~
\ee
Here we used (\ref{Deltabeta}), (\ref{Deltabeta2}), (\ref{genS2}), (\ref{ThetaS}), (\ref{qPoch:an}), (\ref{qrefl}). We next define the operator
\be
\mathcal{Z}=\mathcal{J}^N~,\quad \mathcal{J}=\int_{\i\mathbb{R}} \!\rd \chi \;\mathcal{S}(\chi)~,
\ee
yielding the state
\begin{multline}\label{Zgen}
\mathcal{Z}\ket{\alpha}= \sum_{\ul \ell\in\mathbb{F}^N}\int\!\rd^N \!\ul \chi\; \Delta(\ul \chi,\ul \ell)\prod_j \Delta_0(\chi_j,\ell_j)(w_j)_1^{\sqrt{\beta}\alpha}(w_j)_2^{\sqrt{\beta}\alpha}\;\times\\
\times\bigotimes_{i=1,2}\exp\left(\sum_{n>0}\frac{\sum_j (w_j^n)_i}{q_i^{n/2}-q_i^{-n/2}}\;\a_{-n,i}\right)\e^{\sqrt{\beta}N\Q_i}\ket{\alpha}~.
\end{multline}
Using the algebra representation  (\ref{qviriso}) we can finally match (\ref{ZNlens}), (\ref{ZNS2xS1}), (\ref{ZN:S2xS1A})  identifying 
\be
\begin{array}{|c|c|c|c|}
\hline
&L(r,1)&S^2\times S^1\textrm{ (index) }&S^2\times S^1\textrm{ (twisted index) }\\
\hline
\kappa_2~&~0~&~0~&~0\\
\hline
\kappa_1~&\omega\sqrt{\beta}(\sqrt{\beta}N-Q_\beta+\alpha)&0& 2\sqrt{\beta}\alpha+R(N-1)+2\\
\hline
\kappa_0~&\multicolumn{3}{c|}{1+\frac{\alpha}{2N\sqrt{\beta}}}\\
\hline
\eta_1~&-&-\sqrt{\beta}(\alpha+\sqrt{\beta}N-Q_\beta)&(1-N)\left(\beta-\frac{R}{2}\right)\\
\hline
\end{array}~~~.
\ee

In order to show that the generating function (\ref{Zgen}) satisfies \mbox{$q$-Virasoro} constraints, we have to verify that
\be\label{TS:r}
[\T_{n,i},\mathcal{S}(\chi)]=\textrm{total difference}=\sum_{\ell\in\mathbb{F}}\left(\mathcal{O}_\ell(\lambda_i+\chi)_i-\mathcal{O}_\ell(\chi)_i\right)~,
\ee
for some ($n$-dependent) operator $\mathcal{O}_\ell(\chi)_i$ and $\lambda_i\in\mathbb{C}$. Indeed, the relation (\ref{TSi:1}) can be used to conclude that (\ref{TS:r}) holds true with\footnote{For $L(r,1)$ we require for simplicity that $\sqrt{\beta}\P_i/r$ has integer eigenvalues.}
\be
\mathcal{O}_\ell(X)_1=(w)_2\,{\sf O}(w)_1\otimes \S(w)_2~,\quad
\mathcal{O}_\ell(X)_2=(w)_1\,\S(w)_1\otimes {\sf O}(w)_2~.
\ee
Then two commuting sets of \mbox{$q$-Virasoro} constraints for the generating function (\ref{Zgen}) follow by the usual algebra representation (\ref{qviriso}).

\section{Inclusion of Chern-Simons terms}\label{sec:CS}
In the previous sections we have reviewed how supersymmetric localization allows us to compute partition functions or Wilson loop generating functions of 3d $\mathcal{N}=2$ ${\rm U}(N)$ YM-CS theories on various compact spaces. Focusing on theories with no bare CS level, we have shown that such observables have a natural interpretation in terms of what we called the \mbox{$q$-Virasoro} modular double. Exploiting the free boson representation of this construction, we have derived two commuting sets of (infinitely-many) differential constraints that the generating functions have to satisfy. 

In this section we analyze the inclusion of a bare CS level and its \mbox{$q$-Virasoro} interpretation. The main observation is that CS terms with integer levels can be represented in the matrix models by ``${\rm SL}(2,\mathbb{Z})$-squares" of $\Theta$ functions (see comment around (\ref{Theta1loop}), footnote \ref{thetadefects} and \cite{Beem:2012mb} for more details)
\be
\e^{-S_{\rm CS}}=\prod_{j=1}^N\left(\frac{\Theta(-q_1^{1/2}(w_j)_1;q_1)}{\Theta(-q_1^{1/2};q_1)}\frac{\Theta(-q_2^{1/2}(w_j)_2;q_2)}{\Theta(-q_2^{1/2};q_2)}\right)^{\kappa_2}~,
\ee
where $(w_j)_i$ is a fundamental weight variable introduced in (\ref{table:virgauge}) and $\kappa_2$ is the integer CS level. In fact, by using the ${\rm SL}(2,\mathbb{Z})$ modular properties (\ref{ThetaS}), (\ref{Thetaid}), (\ref{ThetaA}) of the $\Theta$ function and (\ref{qPoch:an}) we have 
\be
\frac{\Theta(-q_1^{1/2}(w)_1;q_1)}{\Theta(-q_1^{1/2};q_1)}\frac{\Theta(-q_2^{1/2}(w)_2;q_2)}{\Theta(-q_2^{1/2};q_2)}=
\left\{\begin{array}{lll}
\e^{-\frac{\i\pi X^2}{\omega_1\omega_2}}&:&S^3_b\\
\e^{-\frac{\i\pi X^2}{r\omega_1\omega_2}}\e^{-\frac{\i\pi \ell^2}{r}}&:&L(r,1)\\
x^\ell&:&S^2\times S^1\textrm{\small (index) }\\
x^\ell&:&S^2\times S^1\textrm{\small (twisted index) }
\end{array}\right.~,~~~
\ee
reproducing the localized CS action on the various backgrounds. From the \mbox{$q$-Virasoro} perspective, the inclusion of CS terms in the gauge theory modifies the external Fock states on which we evaluate the free boson operators by inserting additional vertex operators creating particular coherent states\footnote{The operators $\V_{\pm,i}$ can be thought of as particular specializations of the vertex operator defined in (\ref{Vvertex}) when acting on the (dual) vacuum. While it is possible to use the more general vertex operator (\ref{Vvertex}), we use its specializations in order to avoid unnecessary clutterings.}
\be\label{CSvertex}
\ket{\alpha}\to \mathcal{V}_{-}^{\kappa_2}\ket{\alpha}~,\quad \bra{\alpha}\to\bra{\alpha}\mathcal{V}_{+}^{\kappa_2}~,\quad \mathcal{V}_\pm={\sf V}_{\pm,1}\otimes {\sf V}_{\pm,2}~, \quad {\sf V}_{\pm,i}=\e^{\pm \sum_{n>0}\frac{(-1)^n\lambda_{\pm n,i}}{(q_i^{n/2}-q_i^{-n/2})(t_i^{n/2}-t_i^{-n/2})}}~,
\ee
where we recall the definition (\ref{lambdabasis}) of the operators $\lambda_{n,i}$. We can check that the vertex operators (\ref{CSvertex}) have the desired property by considering their OPE with the \mbox{$q$-Virasoro} screening currents (we drop a normal ordering constant)
\be
{\sf V}_{+,i}\; \S(w)_i\; {\sf V}_{-,i}=
{\sf V}_{-,i}\; \S(w)_i\; {\sf V}_{+,i}\;\Theta(-q_i^{1/2} (w)_i;q_i)~.
\ee
%{\sf V}_{+,i}{\sf V}_{-,i}&={\sf V}_{-,i}{\sf V}_{+,i}\;v_{+-}(q_i,t_i)~,\quad v_{+-}(q_i,t_i)=\frac{(q_i;q_i,t_i,p_i^2)_\infty}{(p_i q_i;q_i,t_i,p_i^2)_\infty}~.
For instance, in the \mbox{$q$-Virasoro} modular double algebra associated to the $S^3_b$ geometry studied in section \ref{sec:S3} we have
\begin{multline}\label{ZNS3CS}
\mathcal{V}_+^{\kappa_2}\mathcal{Z}\mathcal{V}_-^{\kappa_2}\ket{\alpha}= \int\!\rd^N\!\ul  X\;\e^{-\frac{\i\pi\kappa_2}{\omega_1\omega_2}\sum_j X_j^2}\;\mathcal{V}_-^{\kappa_2}\prod_j\mathcal{S}(X_j)\mathcal{V}_+^{\kappa_2}\ket{\alpha}=\\
=\int\!\rd^N\!\ul  X\; \Delta_S(\ul X)\;\e^{-\frac{\i\pi\kappa_2}{\omega_1\omega_2}\sum_j X_j^2}\;\e^{\frac{2\pi\i \omega\sqrt{\beta}}{\omega_1\omega_2}(\sqrt{\beta}N-Q_\beta+\alpha)\sum_j X_j}\; \times\\
\times\bigotimes_{i=1,2}\exp\left(\sum_{n>0}\frac{\sum_j (w_j^n)_i}{q_i^{n/2}-q_i^{-n/2}}\;\a_{-n,i}-\sum_{n>0}\frac{\kappa_2(-1)^n}{(q_i^{n/2}-q_i^{-n/2})(t_i^{n/2}-t_i^{-n/2})}\;\lambda_{-n,i}\right)\e^{\sqrt{\beta}N\Q_i}\ket{\alpha}~,
\end{multline}
where we dropped a proportionality factor. Using the algebra representation (\ref{qviriso}) we finally get the matrix model (\ref{ZNS3}) with $\kappa_2\neq 0$. We also see that while (anti-)fundamental chiral matter can be included by shifting the time variables $\ul\tau_{i}$ in the potential $\sum_j V(X_j|\ul \tau_1,\ul \tau_2)$, the inclusion of CS terms shifts the power sums $\sum_j (w_j^n)_i$. Also, the inclusion of CS terms will modify the differential constraints satisfied by the generating functions, which can be computed by action of the \mbox{$q$-Virasoro} modular double currents on the ``dressed" state $\mathcal{V}_+^{\kappa_2}\mathcal{Z}\mathcal{V}_-^{\kappa_2}\ket{\alpha}$.

\subsection{Decoupling hyper multiplets}
Another way to generate an integer CS term in the gauge theory is to couple a pair of fundamental/anti-fundamental chiral multiplets (in fact, a hyper multiplet) and then letting the physical masses go to infinity. In this limit the multiplets can be integrated out and their contribution to the partition function simplifies dramatically leaving behind an integer CS unit. Focusing on the $S^3_b$ geometry for concreteness, this can be explicitly seen from the 1-loop matter contribution to the matrix model potential 
\be
V(X)_{\rm matter}=-\ln S_2(X+M|\ul \omega)-\ln S_2(-X+\bar M|\ul \omega)~,
\ee 
where $M=-\i M^{\mathbb{R}}+\frac{\omega}{2}\Delta$, $\bar M=-\i \bar M^{\mathbb{R}}+\frac{\omega}{2}\bar\Delta$ are the complexified masses with $M^{\mathbb{R}},\bar M^{\mathbb{R}}$ and $\Delta,\bar \Delta$ being the real masses and Weyl dimensions respectively. We can further split $M^{\mathbb{R}}=M_V+M_A$, $\bar M^{\mathbb{R}}=-M_V+M_A$, where $M_V,M_A$ are the vector and axial masses respectively. Upon specializing to $M_V=0$ and $\Delta=\bar\Delta$, in the decoupling limit $M_A\to \pm\infty$ the matter contribution reduces to 
\be
V(X)_{\rm matter}\sim {\rm sign}(M_A)\frac{\i\pi}{\omega_1\omega_2}X^2~,
\ee
where we neglected divergent background terms and used the asymptotic expansion of the double Sine function
\be\label{asymptote:S2}
\ln S_2(X |\ul\omega)\sim \mbox{sign}({\rm Im}(X)) \frac{\i\pi }{2\omega_1\omega_2}\left(X^2
-\omega X+\frac{\omega^2+\omega_1\omega_2}{6}\right)~, \quad |X|\to\infty~.
\ee
We can now recognize in the above contribution an induced CS level $\kappa_2=-{\rm sign}(M_A)$.

From the \mbox{$q$-Virasoro} viewpoint, the coupling/decoupling procedure of the chiral multiplet pair involves the insertion of the modular double version of the vertex operator (\ref{Hvertex}), which for the $S^3_b$ geometry reads as
\be
\mathcal{H}_{\gamma}(Z)={\sf H}_\gamma(\e^{\frac{2\pi\i}{\omega_1}Z})_1\otimes {\sf H}_\gamma(\e^{\frac{2\pi\i}{\omega_2}Z})_2~,
\ee
and the large momentum limit ${\rm Im}(\omega\beta\gamma)\to\pm\infty$. In order to see that we can take its OPE with the modular double screening current, yielding 
\begin{multline}
\mathcal{H}_{\gamma}(Z)\mathcal{S}(X)=\;:\mathcal{H}_{\gamma}(Z)\mathcal{S}(X):\frac{\e^{\frac{\i\pi}{\omega_1\omega_2}\gamma \beta\omega(X+Z)}}{S_2(\frac{\omega}{2}+\frac{\gamma \beta\omega}{2}+X-Z|\ul\omega)S_2(\frac{\omega}{2}+\frac{\gamma\beta\omega}{2}-X+Z|\ul\omega)}~.
\end{multline}
When acting on the charged Fock vacuum $\ket{\alpha}$ we obtain
\begin{multline}
\mathcal{H}_{\gamma}(Z)\mathcal{S}(X)\ket{\alpha}=\frac{\e^{\frac{\pi\i\omega}{\omega_1\omega_2}Z(\beta\gamma+\sqrt{\beta}\gamma\alpha)}\e^{\frac{2\pi\i\omega}{\omega_1\omega_2}X(1+\sqrt{\beta}\alpha+\frac{\beta\gamma}{2})}}{S_2(\frac{\omega}{2}+\frac{\gamma \beta\omega}{2}+X-Z|\ul\omega)S_2(\frac{\omega}{2}+\frac{\gamma \beta\omega}{2}-X+Z|\ul\omega)}\times\\
\times\bigotimes_{i=1,2}\exp\left(\sum_{n>0}\frac{\e^{\frac{2\pi\i n}{\omega_i}X}}{q_i^{n/2}-q_i^{-n/2}}\;\a_{-n,i}+\sum_{n>0}\frac{(t_i^{\gamma n/2}-t_i^{-\gamma n/2})\e^{\frac{2\pi\i n}{\omega_i}Z}}{(q_i^{n/2}-q_i^{-n/2})(t_i^{n/2}-t_i^{-n/2})}\;\lambda_{-n,i}\right)\e^{\sqrt{\beta}\Q_i(1+\frac{\gamma}{2})}\ket{\alpha}~,
\end{multline}
and upon shifting $\alpha\to\alpha-\sqrt{\beta}\gamma/2$ and taking the limit ${\rm Im}(\omega\beta\gamma)\to\pm\infty$ we get an effective contribution $\kappa_2={\rm sign}({\rm Im}(\omega\beta\gamma))$ to the CS level in the matrix model potential.

It is worth observing that this mechanism is essentially equivalent to the one we have discussed previously: in fact, on the one hand we can write
\be
\mathcal{H}_\gamma(Z)\mathcal{S}(X)=[\mathcal{H}_\gamma(Z)]_0[\mathcal{H}_\gamma(Z)]_-\; \mathcal{S}(X)\;[\mathcal{H}_\gamma(Z)]_+ \; {\rm OPE}(X)~,
\ee
where $[~]_{\pm,0}$ denotes the positive, negative or zero mode part and ${\rm OPE}(X)$ is the normal ordering function giving rise to the CS level in the limit; on the other hand we can split $[\mathcal{H}_\gamma(Z)]_{\pm}=\; [\mathcal{V}_\gamma(Z)\mathcal{V}_{-\gamma}(Z)^{-1}]_\pm$ as in (\ref{Hsplitting}), and in the limit one of the component vertices is predominant over the other and the resulting action on the vacuum is essentially equivalent to the action of $\mathcal{V}_{\pm}$ defined in (\ref{CSvertex}).

\subsection{Pure Chern-Simons and torus knots}\label{pureCS}
Since through the inclusion of the vertex operators (\ref{CSvertex}) we can introduce CS terms in our \mbox{$q$-Virasoro} matrix models, it is interesting to investigate the relation of the latter to pure CS matrix models \cite{Marino:2002fk,Aganagic:2002wv}. Focusing again on the $S^3_b$ geometry, we immediately see that upon setting $\beta=1/2$ the generating function (\ref{ZNS3}) reduces to 
\begin{multline}\label{ZNS:b=1/2}
  Z(\ul\tau_1,\ul\tau_2)=(-4)^{\frac{N(N-1)}{2}}\int_{\i\mathbb{R}^N}\!\rd^N \!\ul X\; 
  \prod_{k<j}\sin\left(\pi \frac{X_k-X_j}{\omega_1}\right) \sin\left(\pi\frac{X_k-X_j}{\omega_2}\right)\times \\
  \times  \e^{\frac{\i \pi \kappa_2}{\omega_1 \omega_2}\sum_j X_j^2}\e^{\frac{2 \pi \i \kappa_1}{\omega_1 \omega_2}\sum_j X_j}
 \prod_{i=1,2} \exp\left(\sum_{n> 0}\tau_{n,i} 
  \sum_j \e^{\frac{2\pi\i n}{\omega_i}X_j}+\tau_{0,i} N\kappa_0 \right) ~,
\end{multline}
corresponding to the generating function of Wilson loops in pure CS theory on $S^3_b$. Physically, the value $\beta=1/2$ corresponds to a massless adjoint chiral multiplet, and hence its 1-loop contribution is trivial due to cancellations between opposite roots.

In contrast to section \ref{sec:limits}, where we considered $\beta=1/2$ as a particular case in the round $S^3$ geometry, here the algebra of the constraints is still given by the \mbox{$q$-Virasoro} modular double. 

%However, it is worth noting that in the case $\beta=1/2$ the \mbox{$q$-Virasoro} algebra has been related \cite{Awata:1996fq} to the ${\rm W}_{1+\infty}^{c=1}$ algebra. 

A particularly interesting situation is when $\omega_1$ and $\omega_2$ are two coprime integers. In this case the matrix model (\ref{ZNS:b=1/2}) corresponds to the Wilson loop generating function for torus knots. Notice that in this limit the deformation parameters $q_1=\e^{2\pi \i\frac{\omega}{\omega_1}}$ and
$q_2=\e^{2\pi \i\frac{\omega}{\omega_2}}$ both go to roots of unity. It is also known that the 
matrix integral (\ref{ZNS:b=1/2}) satisfy usual  Virasoro constraints in this limit \cite{Dubinkin:2013tda}.

\subsection{Refined Chern-Simons}
The \mbox{$q$-Virasoro} matrix model (\ref{qvirgen}), or equivalently the $D^2\times S^1$ generating function (\ref{d2xs1gen}), is of refined CS type \cite{Aganagic:2011sg,Aganagic:2012ne,Aganagic:2012hs}. In fact, the vector and adjoint multiplets provide the Macdonald integration measure $\Delta_\beta(\ul w;q)$ (\ref{Deltabeta}). Considering an $S^1$ fibration over $S^2$ with first Chern class $\kappa$, the partition function of refined CS theory reads as
\be
Z_{\rm rCS}(\kappa)=\int\rd^N \ul W\;\Delta_{\beta_{\textrm{\tiny rCS}}}(\e^{\ul W};q_{\textrm{\tiny rCS}})\;\e^{-\frac{\kappa}{2 g_s}\sum_j W_j^2}~,
\ee
where we used the parametrization $w=\e^W$ and $q_{\textrm{\tiny rCS}}=\e^{g_s}$. In particular, $\kappa=0,1$ corresponds to $S^2\times S^1$ and $S^3$ respectively. Given the relation between the \mbox{$q$-Virasoro} matrix model and refined CS, it is natural to ask whether there is any relation between \mbox{$q$-Virasoro} modular double matrix models, or equivalently 3d compact space generating functions, and refined CS. For instance, the $S^3_b$ partition function (\ref{ZNS3}) with $\kappa_1=0$ reads as 
\be
Z(\ul 0,\ul 0)=\int_{\i\mathbb{R}^N}\!\rd^N\! \ul X\; \Delta_S(\ul X)\;\e^{-\frac{\i\pi\kappa_2}{\omega_1\omega_2}\sum_j X_j^2} ~.
\ee
In order to establish a clear relation with refined CS, let us start by taking the specialization $\beta\omega=b_2\omega_2~,b_2\in\mathbb{Z}_{>0}$, in the $S^3_b$ matrix model, in which case the measure simplifies to 
\begin{multline}
\Delta_S(\ul X)\Big|_{\beta=\frac{b_2\omega_2}{\omega}}=2^{b_2N(N-1)}\prod_{j\neq k}\prod_{n=0}^{b_2-1}\sin\left(\pi\frac{n\omega_2+X_j-X_k}{\omega_1}\right)=\\
=(-2\i^{b_2})^{b_2 N(N-1)}\prod_{j\neq k}\prod_{n=0}^{b_2-1}(\e^{\frac{\i\pi}{\omega_1}(n\omega_2 +X_j-X_k)}-\e^{-\frac{\i\pi}{\omega_1}(n\omega_2+X_j-X_k)})~.
\end{multline}
On the other hand, the Macdonald measure for $\beta_{\textrm{\tiny rCS}}\in\mathbb{Z}_{>0}$ (which is a common specialization in refined CS) simplifies to 
\begin{multline}
\Delta_{\beta_{\textrm{\tiny rCS}}}(\e^{\ul W};q_{\textrm{\tiny rCS}})\Big|_{\beta_{\textrm{\tiny rCS}}\in\mathbb{Z}_{>0}}=\prod_{k\neq j}\prod_{n=0}^{\beta_{\textrm{\tiny rCS}}-1}(1-q_{\textrm{\tiny rCS}}^n \e^{W_j-W_k})=\\
=q_{\textrm{\tiny rCS}}^{N(N-1)\frac{\beta_{\textrm{\tiny rCS}}(\beta_{\textrm{\tiny rCS}}-1)}{4}}\prod_{j\neq k}\prod_{n=0}^{\beta_{\textrm{\tiny rCS}}-1}(q_{\textrm{\tiny rCS}}^{\frac{n}{2}}\e^{\frac{W_j-W_k}{2}}-q_{\textrm{\tiny rCS}}^{-\frac{n}{2}}\e^{\frac{W_k-W_j}{2}})~,
\end{multline}
implying that the $S^3_b$ matrix model collapses to the refined CS matrix model upon identifying $g_s=2\pi\i\omega_2/\omega_1$, $\beta_{\textrm{\tiny rCS}}=b_2$, $W=2\pi\i X/\omega_1$ and $\kappa=\kappa_2$. Similarly, if we take the more general specialization  $\beta\omega=b_1\omega_1+b_2\omega_2$, $b_{1,2}\in\mathbb{Z}_{>0}$, the $\Delta_S(\ul X)$ measure collapses to two copies of the Macdonald measure due to (\ref{quasiper}). We can in fact relax any specialization of the parameters and consider instead the limit where the $S^3_b$ is very squashed, i.e. $|\omega_1/\omega_2|\gg 1$, in which case the double Sine function has the semiclassical behaviour (assuming ${\rm Im}(\omega_2/\omega_1)>0$)
\be
S_2(X|\ul\omega)=\e^{\frac{\i\pi}{2}B_{22}(X|\ul\omega)}(\e^{\frac{2\pi\i}{\omega_1}X};\e^{2\pi\i\frac{\omega_2}{\omega_1}})_\infty(1+O(\e^{2\pi\i\frac{\omega_1}{\omega_2}}))~,
\ee 
and hence
\be
\Delta_S(\ul X)=\Delta_\beta(\e^{\frac{2\pi\i}{\omega_1}\ul X};\e^{2\pi\i\frac{\omega_2}{\omega_1}})(1+O(\e^{2\pi\i\frac{\omega_1}{\omega_2}}))~,
\ee
up to proportionality factors. Therefore, the \mbox{$q$-Virasoro} modular double might give rise to a doubled or non-perturbative version of refined CS. Moreover, it is known that the (large $N$ limit of) refined CS observables are captured by refined (closed) open topological strings \cite{Aganagic:2011sg,Aganagic:2012hs,Iqbal:2011kq}, and thus one can expect that the \mbox{$q$-Virasoro} modular double might also play a role in the non-perturbative description of refined topological strings. This observation is in line with those of \cite{Pasquetti:2011fj,Lockhart:2012vp}, and we will comment more on this aspect in section \ref{sec:summary}.

\section{Generalization to quiver gauge theories}\label{sec:generalizations}
Our discussion on the \mbox{$q$-Virasoro} structures in 3d $\mathcal{N}=2$ YM-CS theories has so far focused on a single node ${\rm U}(N)$ gauge group coupled to 1 adjoint and possibly (anti-)fundamental chirals. The goal of this section is to show that our construction admits a generalization to a huge class of 3d $\mathcal{N}=2$ unitary quiver gauge theories and ${\rm W}_{q,t}(\Gamma)$ algebras of \cite{Kimura:2015rgi}. 

\begin{figure}[!ht]
\leavevmode
\begin{center}
\includegraphics[height=0.15\textheight]{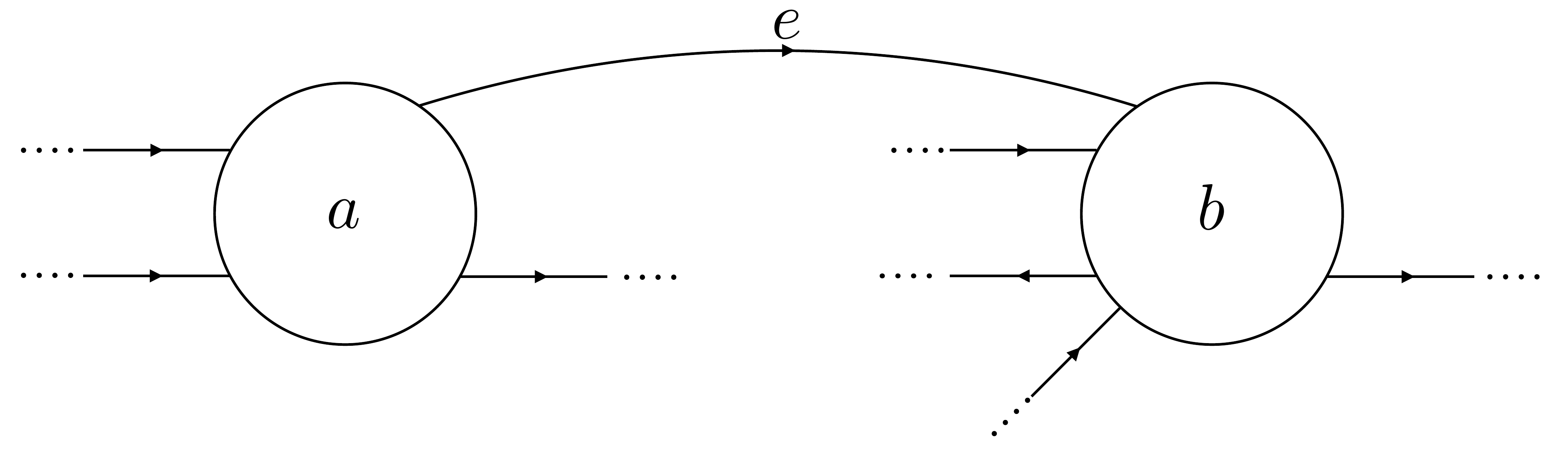}
\end{center}
\caption{Portion of a quiver $\Gamma$. We explicitly displayed 2 nodes $a,b\in\Gamma_0$, an arrow $e\in\Gamma_1$ from $a$ to $b$, and several arrows with source or target in $a$ or $b$.}
\label{Gamma}
\end{figure}

Let us start by recalling some algebraic definition from \cite{Kimura:2015rgi}. A quiver $\Gamma$ is a collection of nodes $\Gamma_0$ and arrows $\Gamma_1$, see figure \ref{Gamma} for an example.
Given two nodes $a,b\in\Gamma_0$ and an arrow $\Gamma_1\ni e:a\to b$, we can associate to the quiver the deformed Cartan matrix $C_{ab}\in |\Gamma_0|\times |\Gamma_0|$
\be
C_{ab}=(1+p^{-1})\delta_{ab}-\sum_{e:b\to a}m_e^{-1}-p^{-1}\sum_{e:a\to b}m_e~,
\ee
the Heisenberg algebra (we display non-vanishing commutators only)
\be
\!\!\!\![\a^a_{n},\a^b_{m}]=\frac{1}{n}(q^\frac{n}{2}-q^{-\frac{n}{2}})(t^\frac{n}{2}-t^{-\frac{n}{2}})p^{\frac{n}{2}}C_{ab}^{[n]}\delta_{n+m,0}~,\quad \!\! [\P^a,\Q^b]=C_{ab}^{[0]}~,\!\! \quad n,m\in\mathbb{Z}\backslash\{0\}~,
\ee
and the screening current
\be
\S^a(w)=\; :\e^{-\sum_{n\neq 0}\frac{w^{-n}}{q^{n/2}-q^{-n/2}}\a_{n}^a}:\e^{\sqrt{\beta}\Q^a}w^{\sqrt{\beta}\P^a}~,
\ee
where $q,t,p=qt^{-1},m_e\in\mathbb{C}$\footnote{To compare with \cite{Kimura:2015rgi} we have to set $(q)_{\rm here}=(q_2)_{\rm there}$, $(t)_{\rm here}=(q_1^{-1})_{\rm there}$.}, while the ${}^{[n]}$ operation means replacing each parameter with its $n^{\rm th}$ power, for instance
\be
C_{ab}^{[n]}=(1+p^{-n})\delta_{ab}-\sum_{e:b\to a}m_e^{-n}-p^{-n}\sum_{e:a\to b}m_e^n~.
\ee
With these data  the ${\rm W}_{q,t}(\Gamma)$ algebra can be defined to be the non-commutative associative algebra generated by the currents $\{\T^a(z)=\sum_{n\in\mathbb{Z}}\T_{n}^a\; z^{-n},a\in\Gamma_0\}$ and given as the commutant up to total differences of the screening currents in the Heisenberg algebra 
\be\label{TSW}
[\T^a_n,\S^b(w)]=\textrm{total difference}~.
\ee
For instance, the single node quiver $\Gamma_0=\{1\}$, $\Gamma_1=\{\emptyset\}$ corresponding to the $A_1$ Lie algebra diagram gives rise to the $q\textrm{-Virasoro}={\rm W}_{q,t}(A_1)$ algebra reviewed in \mbox{section \ref{sec:free_field}}, whereas the $n$-node quiver $\Gamma_0=\{1,\ldots,n\}$, $\Gamma_1=\{e_a:a\to a+1,a=1,\ldots, n-1\}$ corresponding to the $A_n$ Lie algebra gives rise to the ${\rm W}_{q,t}(A_n)$ algebra of \cite{Awata:1995zk}. More generally, for quivers associated to simple Lie algebras the construction of \cite{Kimura:2015rgi} agrees with \cite{1997q.alg.....8006F}.
%The ${\rm W}_{q,t}(\Gamma)$ algebras have been introduce in the context of 5d $\mathcal{N}=1$ gauge theories on $\mathbb{R}^4_{\epsilon_1,\epsilon_2}\times S^1$. The algebra/gauge theory dictionary is the following
%\be
%\begin{array}{c|c}
%\textrm{Algebra}&\textrm{5d gauge theory}\\
%\hline
%\Gamma_0~&~ {\rm U}(N) \textrm{ gauge nodes}\\
%\Gamma_1\ni e:a\to b ~&~ {\rm U}(N_a)\times {\rm U}(N_b) \textrm{ bi-fundamental hyper} \\
%q~,t~,\beta~&~ \e^{\epsilon_2}~,\e^{-\epsilon_1}~,-\epsilon_1/\epsilon_2\\
%m_e~&~\textrm{bi-fundamental mass}
%\end{array}\quad~.
%\ee
%The 5d Nekrasov partition function can be computed by localization in T-equivariant K-theory of the instanton moduli space. The points of the the fixed point set $M^{\rm T}$ under the T-action are labelled by Young diagrams, while the shifted Coulomb branch parameters can be collected into the set
%\be
%\chi=\bigcup_{a\in\Gamma_0}\chi_a~,\quad \chi_a=\{x_{a,n,i}=A_{a,n} t^{1-i}q^{Y^a_{n,i}},n\in[1,\ldots, N_a],i\in[1,\ldots,\infty]\}~.
%\ee
%It was shown in \cite{} that the 5d Nekrasov partition function is computed by the correlator $\braket{0}{\Z}$, where
%\be
%\ket{Z}=\sum_{\chi\in M^{\rm T}}\prod^{\succ}_{x\in \chi}\S^{\i(x)}(x)\ket{0}~,
%\ee
%the map $\i:\chi\to\Gamma_0$ being the node label so that $\i(x)=a$ if $x\in\chi_a$.
%
%Here  ${\sf J}^{\i(x_\emptyset)}_{x_\emptyset}$ is the screening charge
%\be
%{\sf J}^{a}_{x_\emptyset}=\sum_{n\in\mathbb{Z}}\S^{a}(q^n x_\emptyset)~,
%\ee
%$(x_{a,n,i})_\emptyset=A_{a,n} t^{1-i}$ is the ground configuration.

\begin{figure}[!ht]
\leavevmode
\begin{center}
\includegraphics[height=0.19\textheight]{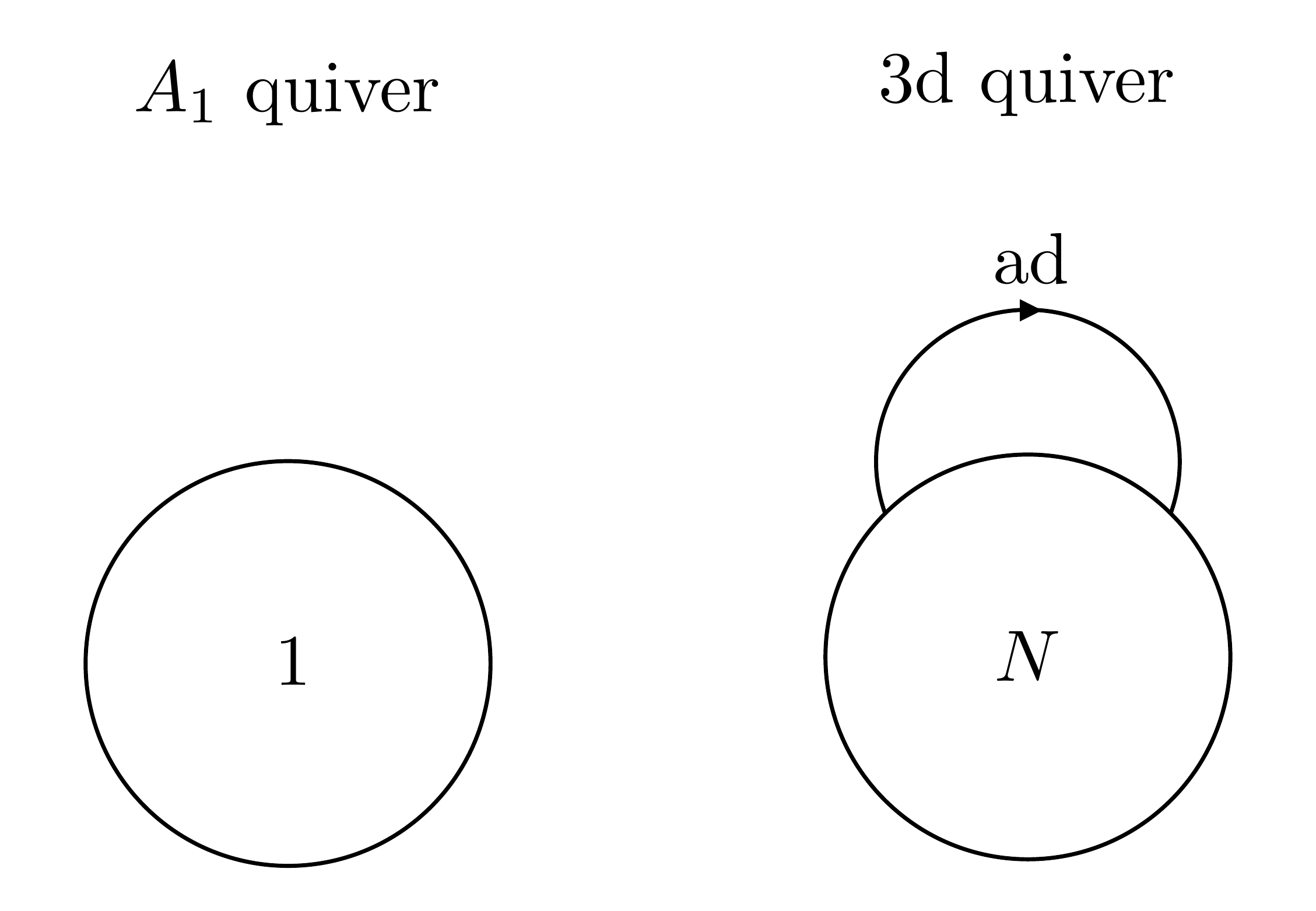}
\end{center}
\caption{The quivers of \mbox{$q$-Virasoro} (left) and the corresponding 3d gauge theory (right).}
\label{A1-3d}
\end{figure}

Following the discussion of subsection \ref{D2xS1}, we can now associate to the ${\rm W}_{q,t}(\Gamma)$ algebra a 3d $\mathcal{N}=2$ unitary quiver gauge theory on $D^2\times S^1$, whose Wilson loop generating function will be reproduced by the action of the ${\rm W}_{q,t}(\Gamma)$ screening charges on a charged Fock vacuum $\ket{\ul\alpha}$, $\ul\alpha=\{\alpha_a,a\in \Gamma_0\}$, namely
\be
Z(\{\ul\tau^a\})\simeq \oint\prod_{a=1}^{|\Gamma_0|}\frac{\rd^{N_a} \ul w_a}{2\pi\i\ul w_a}\prod_{a=1}^{|\Gamma_0|}\prod_{j=1}^{N_a}\S^{a}(w_{a,j})\ket{\ul\alpha}~.
\ee
It is important at this point to not confuse the $\Gamma$ quiver of the algebra with the unitary quiver of the 3d theory. For instance, the $A_1$ quiver associated to the \mbox{$q$-Virasoro} algebra has a single node and no arrows at all, while the dual gauge theory involves a ${\rm U}(N)$ gauge vector and 1 adjoint chiral multiplet, whose quiver description usually consists of a round node for the gauge group and a loop arrow for the adjoint, as depicted in figure \ref{A1-3d}. 
In order to determine the dual 3d gauge theory description for the general case, the key point is to understand the measure arising from the product of several screening currents
\begin{multline}\label{SSS:quiver}
\prod_{a=1}^{|\Gamma_0|}\prod_{j=1}^{N_a}\S^{a}(w_{a,j})=\; :\prod_{a=1}^{|\Gamma_0|}\prod_{j=1}^{N_a}\S^{a}(w_{a,j}):\prod_{a=1}^{|\Gamma_0|}c_\beta(w_{a},1;q)\Delta_\beta(w_{a};q)\prod_{j=1}^{N_a} w_{a,j}^{\beta(N_a-1)}\times\\
\times\prod_{a=1}^{|\Gamma_0|}\prod_{e:a\to a}\frac{1}{c_\beta(w_{a},m_e;q)}\prod_{1\leq j\neq k\leq N_a}\frac{(t m_e w_{a,k}w_{a,j}^{-1};q)_\infty}{(m_e w_{a,k}w_{a,j}^{-1};q)_\infty} \prod_{j=1}^{N_a}w_{a,j}^{-\beta(N_a-1)}\times\\
\times\prod_{1\leq a< b\leq |\Gamma_0|}\prod_{j=1}^{N_a}\prod_{k=1}^{N_b}\prod_{e:a\to b}\frac{ (t m_e w_{b,k} w_{a,j}^{-1};q)_\infty}{(m_e w_{b,k} w_{a,j}^{-1};q)_\infty}\;w_{a,j}^{-\beta}\prod_{e:b\to a}\frac{(q m_e^{-1} w_{b,k} w_{a,j}^{-1};q)_\infty}{(q t^{-1} m_e^{-1} w_{b,k} w_{a,j}^{-1};q)_\infty}\;w_{a,j}^{-\beta}~.
\end{multline}
From this expression we can immediately read off the corresponding 3d $\mathcal{N}=2$ gauge theory: it is a $\Gamma$ quiver YM theory with ${\rm U}(N_a)$ gauge nodes each coupled to 1 adjoint chiral multiplet, 1 bi-fundamental hyper multiplet (actually a pair of fundamental/anti-fundamental chirals) for each arrow connecting different gauge nodes and 1 adjoint hyper multiplet (a pair of adjoint chirals) for each loop edge. The generating function of the theory is identified with a heighest weight state of the ${\rm W}_{q,t}(\Gamma)$ algebra and will satisfy the associated constraints
\be
T^a(z|\ul\tau^a)Z(\{\ul\tau^a\})={\rm Pol}(z)\quad \Rightarrow\quad T_n^a(\ul\tau^a)Z(\{\ul\tau^a\})=0~,\quad n>0
\ee
by construction. (Anti-)fundamental chiral multiplets or CS levels for each gauge node can be added on top of this construction as insertion of addition vertex operators, and may be represented by auxiliary square nodes or integer labels respectively. 

\begin{figure}[!ht]
\leavevmode
\begin{center}
\includegraphics[height=0.20\textheight]{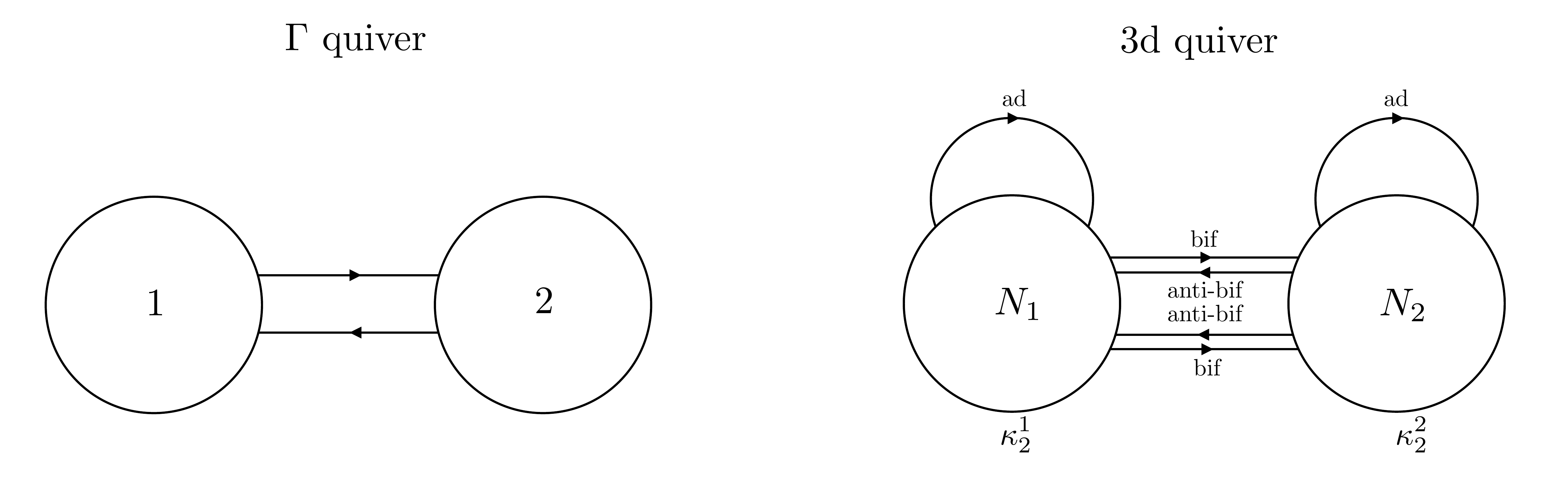}
\end{center}
\caption{The ``ABJ(M) quiver'' of the algebra (left) and the gauge theory (right).}
\label{ABJquiver}
\end{figure}

As it should be clear from the previous sections, if we want to discuss 3d theories on compact spaces we should construct the modular double of the ${\rm W}_{q,t}(\Gamma)$ algebras. In the $q\textrm{-Virasoro}={\rm W}_{q,t}(A_1)$ case, our construction of the modular double only relied on the property (\ref{TSW}) of the screening currents, and therefore we can generalize our analysis to arbitrary ${\rm W}_{q,t}(\Gamma)$ algebras following the recipe given in sections \ref{sec:S3}, \ref{sec:other} for the various geometries.  As an application, in the following we will consider the simple but important example of  the (mass deformed) ABJ(M) theory on $S^3_b$, where the relevant quivers are shown in figure \ref{ABJquiver}.

\subsection{ABJ(M) theory}
The ABJ theory \cite{Aharony:2008gk} is the $\mathcal{N}=6$ ${\rm U}(N_1)_{\kappa_2}\times {\rm U}(N_2)_{-\kappa_2}$ CS theory with 1 bi-fundamental and 1 anti-bi-fundamental hyper multiplets, where the subindex denotes the CS level. In the case $N_1=N_2$ the theory specializes to the ABJM model \cite{Aharony:2008ug}. Using the notation of section \ref{sec:S3}, its $S^3_b$ partition function reads as (see for instance \cite{Hatsuda:2016uqa})
\begin{multline}\label{ZS3ABJ}
Z_{\rm ABJ}=\mathcal{N}_0\int_{\i\mathbb{R}}\prod_{a=1,2}\!\rd^{N_a} \!\ul X_a\;\e^{-\frac{\i\pi\kappa_2}{\omega_1\omega_2}(\sum_{j=1}^{N_1} X_{1,j}^2-\sum_{k=1}^{N_2} X_{2,k}^2)}\times\\ 
\times\prod_{a=1,2}\prod_{1\leq j<k\leq N_a}(2\i)^2\sin\left(\pi\frac{X_{a,j}-X_{a,k}}{\omega_1}\right)\sin\left(\pi\frac{X_{a,j}-X_{a,k}}{\omega_2}\right)\times\\
\times\prod_{j=1}^{N_1}\prod_{k=1}^{N_2}\frac{S_2\left(\frac{\omega}{2}+X_{2,k}-X_{1,j}+\frac{\omega}{4}|\ul\omega\right)^2}{S_2\left(\frac{\omega}{2}+X_{2,k}-X_{1,j}-\frac{\omega}{4}|\ul\omega\right)^2}~.
\end{multline}

In order to describe the partition function (or the generating function) of the ABJ theory through ${\rm W}_{q,t}(\Gamma)$ techniques, let us start by considering a two node quiver $\Gamma$ with two oppositely oriented\footnote{The orientation does not actually matter for the ABJ(M) theory.} arrows connecting the two nodes as in figure \ref{ABJquiver}. The corresponding product of screening currents is a simple specialization of (\ref{SSS:quiver}) 
\begin{multline}
\prod_{a=1}^{2}\prod_{j=1}^{N_a}\S^{a}(w_{a,j})=\; :\prod_{a=1}^{2}\prod_{j=1}^{N_a}\S^{a}(w_{a,j}):\Delta_\beta(\ul w_{a};q)c_\beta(\ul w_{a};q)\prod_{a=1}^{2}\prod_{j=1}^{N_a} w_{a,j}^{\beta(N_a-1)}\times\\
\prod_{j=1}^{N_1}w_{1,j}^{-2\beta N_2}\prod_{k=1}^{N_2}\frac{ (t m_{12} w_{2,k} w_{1,j}^{-1};q)_\infty}{(m_{12} w_{2,k} w_{1,j}^{-1};q)_\infty}\frac{(q m_{21}^{-1} w_{2,k} w_{1,j}^{-1};q)_\infty}{(q t^{-1} m_{21}^{-1} w_{2,k} w_{1,j}^{-1};q)_\infty}~,
\end{multline}
where we set $m_e=m_{12}$ for the arrow $e:1\to 2$ and $m_e=m_{21}$ for the arrow $e:2\to 1$.
In order to describe the theory on $S^3_b$ we consider the modular double construction of section \ref{sec:S3}, namely we define
\be
\begin{split}
\mathcal{S}^a(X_{a,j})&=(w_{a,j})_1(w_{a,j})_2\; \S^a(w_{a,j})_1\otimes\S^a(w_{a,j})_2~,\\
(w_{a,j})_i&=\e^{\frac{2\pi\i}{\omega_i}X_{a,j}}~,\quad (m_e)_i=\e^{\frac{2\pi\i}{\omega_i}M_e}~,\quad e\in\{(12),(21)\}~,\quad i=1,2~,
\end{split}
\ee
with the ${\rm SL}(2,\mathbb{Z})$ gluing as in table (\ref{Sgluing}). Here $M_e$ are interpreted as the complexified masses for the bi-fundamental hypers, namely $\i M_e=M_e^{\mathbb{R}}+\i\frac{\omega}{2}\Delta$ where $\Delta$ is the Weyl dimension, which we take to be $\Delta=1/2$. The corresponding product of screening charges yields the operator
\begin{multline}
\mathcal{Z}=\int_{\i\mathbb{R}}\prod_{a=1,2}\!\rd^{N_a} \!\ul X_a\;\prod_{a=1}^{2}\prod_{j=1}^{N_a}\mathcal{S}^{a}(X_{a,j})=\\
=\e^{-\frac{\i\pi\omega\beta N_1 N_2}{\omega_1\omega_2}(M_{12}-M_{21})}\int_{\i\mathbb{R}}\prod_{a=1,2}\!\rd^{N_a} \!\ul X_a\; :\prod_{a=1}^{2}\prod_{j=1}^{N_a}\mathcal{S}^{a}(X_{a,j}):\times\\
\times \prod_{a=1,2}\e^{\frac{2\pi\i\omega\sqrt{\beta}}{\omega_1\omega_2}((-1)^a\sqrt{\beta}(N_2-N_1)- Q_\beta)\sum_{j=1}^{N_a} X_{a,j}}\;\Delta_S(\ul X_a)\times\\
\times\prod_{j=1}^{N_1}\prod_{k=1}^{N_2}\frac{S_2(\omega-M_{21}+X_{2,k}-X_{1,j}|\ul\omega)S_2(\beta\omega+M_{12}+X_{2,k}-X_{1,j}|\ul\omega)}{S_2(\omega-\omega\beta-M_{21}+X_{2,k}-X_{1,j}|\ul\omega)S_2(M_{12}+X_{2,k}-X_{1,j}|\ul\omega)}~.
\end{multline}
We now include CS terms as discussed in section \ref{sec:CS}. The ${\rm W}_{q,t}(\Gamma)$ generalization of the vertex operators (\ref{CSvertex}) is given by
\be
\mathcal{V}^a_\pm={\sf V}^a_{\pm,1}\otimes {\sf V}^a_{\pm,2}~, \quad {\sf V}^a_{\pm,i}=\e^{\pm \sum_{n>0}\frac{(-1)^n\lambda^a_{\pm n,i}}{(q_i^{n/2}-q_i^{-n/2})(t_i^{n/2}-t_i^{-n/2})}}~,
\ee
where we have introduced the basis 
\be
\lambda^a_{n,i}=\a^b_{n,i} (C_i^{[-n]})^{-1}_{ba}p_i^{\frac{n}{2}}~,\quad \P^a_{\lambda,i}=\P^b_i (C^{[0]}_i)^{-1}_{ba}~,\quad \Q_{\lambda,i}^a=\Q^b_i (C^{[0]}_i)^{-1}_{ba}~, \quad n\in\mathbb{Z}\backslash\{0\}~~~~~
\ee
of the two commuting ($i=1,2$) Heisenberg algebras satisfying (we display non-trivial relations only)
\be
[\a^a_{n,i},\lambda^b_{m,i}]=\frac{1}{n}(q_i^\frac{n}{2}-q_i^{-\frac{n}{2}})(t_i^\frac{n}{2}-t_i^{-\frac{n}{2}})\delta_{a,b}\delta_{n+m,0}~,\quad\!\! [\P^a_i,\Q^b_{\lambda,i}]=[\P^a_{\lambda,i},\Q^b_i]=\delta_{a,b}~,\quad \!\! n,m\in\mathbb{Z}\backslash \{0\}~.
\ee
We can now consider the dressed operator
\begin{multline}
\prod_{a=1,2}\left(\mathcal{V}_+^a\right)^{\kappa_2^a}\mathcal{Z}\prod_{a=1,2}\left(\mathcal{V}_-^a\right)^{\kappa_2^a}=\\
=\e^{-\frac{\i\pi\omega\beta N_1 N_2}{\omega_1\omega_2}(M_{12}-M_{21})}\int_{\i\mathbb{R}}\prod_{a=1,2}\!\rd^{N_a}\!\ul  X_a\; :\prod_{a=1}^{2}\left(\mathcal{V}_+^a\right)^{\kappa_2^a}\left(\prod_{j=1}^{N_a}\mathcal{S}^{a}(X_{a,j})\right)\left(\mathcal{V}_-^a\right)^{\kappa_2^a}:\times\\
\times \prod_{a=1,2}\e^{-\frac{\i\pi\kappa_2^a}{\omega_1\omega_2}\sum_{j=1}^{N_a} X_{a,j}^2} \e^{\frac{2\pi\i\omega\sqrt{\beta}}{\omega_1\omega_2}((-1)^a\sqrt{\beta}(N_2-N_1)- Q_\beta)\sum_{j=1}^{N_a} X_{a,j}}\;\Delta_S(\ul X_a)\times\\
\times\prod_{j=1}^{N_1}\prod_{k=1}^{N_2}\frac{S_2(\omega-M_{21}+X_{2,k}-X_{1,j}|\ul\omega)S_2(\beta\omega+M_{12}+X_{2,k}-X_{1,j}|\ul\omega)}{S_2(\omega-\omega\beta-M_{21}+X_{2,k}-X_{1,j}|\ul\omega)S_2(M_{12}+X_{2,k}-X_{1,j}|\ul\omega)}~,
\end{multline}
where the equality holds up to constant proportionality factors. The action of this operator on the charged Fock vacuum $\ket{\ul\alpha}$ defined by
\be
\ul\alpha=\{\alpha_a,a=1,2\}~,\quad \ket{\ul\alpha}=\otimes_{i=1,2} \e^{\sum_a\alpha_a\Q_{\lambda,i}^a}\ket{0}~,\quad \a^a_{n>0}\ket{0}=0~,\quad \P^a_i\ket{\ul\alpha}=\alpha_a\ket{\ul\alpha}~,~~~
\ee
yields the state
\begin{multline}
\prod_{a=1,2}\left(\mathcal{V}_+^a\right)^{\kappa_2^a}\mathcal{Z}\prod_{a=1,2}\left(\mathcal{V}_-^a\right)^{\kappa_2^a}\ket{\ul\alpha}=\\
=\e^{-\frac{\i\pi\omega\beta N_1 N_2}{\omega_1\omega_2}(M_{12}-M_{21})}\int_{\i\mathbb{R}}\prod_{a=1,2}\!\rd^{N_a}\!\ul  X_a\; \prod_{a=1,2}\e^{-\frac{\i\pi\kappa_2^a}{\omega_1\omega_2}\sum_{j=1}^{N_a} X_{a,j}^2} \e^{\frac{2\pi\i\kappa_1^a}{\omega_1\omega_2}\sum_{j=1}^{N_a} X_{a,j}}\times\\
\times\Delta_S(\ul X_a)\prod_{j=1}^{N_1}\prod_{k=1}^{N_2}\frac{S_2(\omega-M_{21}+X_{2,k}-X_{1,j}|\ul\omega)S_2(\beta\omega+M_{12}+X_{2,k}-X_{1,j}|\ul\omega)}{S_2(\omega-\omega\beta-M_{21}+X_{2,k}-X_{1,j}|\ul\omega)S_2(M_{12}+X_{2,k}-X_{1,j}|\ul\omega)}\times\\
\times\bigotimes_{i=1,2}\exp\left(\sum_{n>0}\frac{\sum_{j=1}^{N_a}(w_{a,j}^n)_i}{q_i^{n/2}-q_i^{-n/2}}\;\a^a_{-n,i}- \sum_{n>0}\frac{\kappa_{2}^a(-1)^n}{(q_i^{n/2}-q_i^{-n/2})(t_i^{n/2}-t_i^{-n/2})}\;\lambda^a_{-n,i}\right)\e^{\sqrt{\beta}N_a\Q^a_{i}}\ket{\ul\alpha}~,
\end{multline}
up to proportionality factors, with $\kappa_1^a=\omega\sqrt{\beta}((-1)^a\sqrt{\beta}(N_2-N_1)- Q_\beta+\alpha_a)$. Using the representation
\be
\a^a_{-n,i}\simeq (q^{\frac{n}{2}}_i-q^{-\frac{n}{2}}_i)\tau_{n,i}^a~,\quad \lambda^a_{n,i}\simeq \frac{1}{n}(t_i^\frac{n}{2}-t_i^{-\frac{n}{2}})\frac{\partial}{\partial \tau^b_{n,i}}~,\quad \Q^a_{\lambda,i}\simeq \tau_{0,i}^a~,\quad \P^a_i=\frac{\partial}{\partial \tau^a_{0,i}}~,\quad n>0~,
\ee
this state describes the generating function of the mass deformed ABJ theory coupled to two additional adjoint chiral multiplets and FI parameters
\begin{multline}
\prod_{a=1,2}\left(\mathcal{V}_+^a\right)^{\kappa_2^a}\mathcal{Z}\prod_{a=1,2}\left(\mathcal{V}_-^a\right)^{\kappa_2^a}\ket{\ul\alpha}\simeq Z_{\rm ABJ}(\beta,\{M_e\},\{\kappa^a_1\}|\{\ul\tau^a_1,\ul\tau^a_2\})=\\
=\langle \prod_{i=1,2}\exp\left(\sum_{a,b=1}^2\sum_{n>0} \tau_{n,i}^b\left(\delta_{ba}\sum_{j=1}^{N_a}(w_{a,j}^n)_i- (C^{[n]}_i)_{ba}^{-1}p_i^{-n/2}\frac{(-1)^n \kappa_{2}^a}{t_i^{n/2}-t_i^{-n/2}}\right)\right) \rangle~,
\end{multline}
where
\be
\begin{split}
\langle~1~\rangle&=\mathcal{N}_0\int_{\i\mathbb{R}}\prod_{a=1,2}\!\rd^{N_a} \!\ul X_a\;\prod_{a=1,2}\e^{-\frac{\i\pi\kappa_2^a}{\omega_1\omega_2}\sum_{j=1}^{N_a} X_{a,j}^2} \e^{\frac{2\pi\i\kappa_1^a}{\omega_1\omega_2}\sum_{j=1}^{N_a} X_{a,j}}\;\Delta_S(\ul X_a)\times\\
&\quad\times\prod_{j=1}^{N_1}\prod_{k=1}^{N_2}\frac{S_2(\omega-M_{21}+X_{2,k}-X_{1,j}|\ul\omega)S_2(\beta\omega+M_{12}+X_{2,k}-X_{1,j}|\ul\omega)}{S_2(\omega-\omega\beta-M_{21}+X_{2,k}-X_{1,j}|\ul\omega)S_2(M_{12}+X_{2,k}-X_{1,j}|\ul\omega)}~,\\
\mathcal{N}_0&=\e^{-\frac{\i\pi\omega\beta N_1 N_2}{\omega_1\omega_2}(M_{12}-M_{21})}\prod_{i=1,2}\e^{\sum_{a,b=1}^2 \tau_{0,i}^b(\delta_{ab}\sqrt{\beta}N_a+(C^{[0]}_i)^{-1}_{ba}\alpha_a)}~.
\end{split}
\ee
We can now specialize to $\beta=1/2$, which effectively removes the adjoint chiral multiplets and reduces $\Delta_S(\ul X_a)$ to pure vector contributions (see subsection \ref{pureCS})
\be
\Delta_S(\ul X_a)\Big|_{\beta=\frac{1}{2}}=\prod_{j,k=1}^{N_a}S_2(X_{a,j}-X_{a,k}|\ul\omega)=\prod_{i=1,2}\prod_{1\leq j<k\leq N_a}2\i\sin\left(\pi\frac{X_{a,j}-X_{a,k}}{\omega_i}\right)~.
\ee
Moreover, if we set
\be\label{ABJ:parameters}
\kappa_2^1=-\kappa_2^2=\kappa_2~,\quad \kappa_1^1=\kappa_1^2=0~,\quad M_{12}=M_{21}=\frac{\omega}{4}~,
\ee
the CS levels have opposite signs, the FI parameters are set to zero and the bi-fundamentals are massless, and hence we are effectively describing the ABJ theory. This ${\rm W}_{q,t}(\Gamma)$ modular double description would allow us to write explicitly the Ward identities satisfied by ABJ generating function by acting with the algebra generators.

\section{Summary, comments and outlook}\label{sec:summary}
In this work we have shown that a wide class of 3d $\mathcal{N}=2$ unitary quiver gauge theories on compact spaces hides a modular double ${\rm W}_{q,t}$ symmetry, which we defined. Our argument was based on the realization of supersymmetric Wilson loop generating functions as Fock states obtained through the action of vertex operators and  screening charges of the modular double on a vacuum state. Our interpretation implies the existence of two ${\rm SL}(2,\mathbb{Z})$-related commuting sets of ${\rm W}_{q,t}$ constraints (Ward identities) annihilating the YM generating functions, corresponding to highest weight conditions. As recently stressed in \cite{Awata:2016riz}, these type of deformed Ward identities may be regarded as a further step towards a proper definition of $q$-CFT theories, a deformation of ordinary 2d CFTs. While most of the studies have so far focused on the chiral description, our work shows that it is possible to consistently couple different chiral sectors into well-defined modular invariant objects \cite{Nieri:2013yra,Nieri:2013vba}. This is familiar in 2d CFTs, where the invariance w.r.t the Moore-Seiberg groupoid puts severe constraints on the physical theories \cite{Moore:1988qv}. From this perspective it is not totally surprising that the structure of the modular double is dictated by the compact space geometries: in fact, localization on spaces with boundaries (see \cite{Sugishita:2013jca,Yoshida:2014ssa,Faizal:2016skd,Aprile:2016gvn} for recent discussions) is notoriously more complicated than on closed spaces where there are no ambiguities due to the boundary (indeed, most of the gauge theory dualities have been tested by using compact space observables).

There are a number of further directions worth to be studied, physically and mathematically. First of all, the ${\rm W}_{q,t}$ modular double symmetry of 3d $\mathcal{N}=2$ quiver gauge theories on compact spaces that we have considered represents a new tool for studying these theories, which should supplement the existing large $N$ \cite{Drukker:2010nc,Drukker:2011zy} or Fermi gas techniques \cite{Marino:2011eh,Codesido:2014oua,Grassi:2014uua,Assel:2015hsa}. Moreover, it is very likely that the 3d dualities mentioned in the introduction (see also \cite{Benvenuti:2016wet} for recent new results) have a natural and simple interpretation in ${\rm W}_{q,t}$ language, perhaps along the lines of \cite{Gomis:2014eya} in the more familiar 4d AGT context. It is also interesting to observe the appearance of ${\rm W}$-like symmetries in these theories, which are usually associated to area-preserving diffeomorphisms of membranes (a possible connection between ${\rm W}_{q,t}$ algebras and the physics of membranes was already pointed out in \cite{Awata:1996fq}). 

Secondly, the 5d gauge theory origin of the quiver ${\rm W}_{q,t}$ algebras immediately rises the question whether our 3d gauge theory inspired construction of the modular double can describe the parent 5d theories on compact spaces as well. While in the chiral case (i.e. 5d theories on $\mathbb{R}^4\times S^1$) the answer is clearly affirmative according to the results of \cite{Kimura:2015rgi,Aganagic:2013tta,Aganagic:2014oia,Aganagic:2015cta}, in the non-chiral case (i.e. 5d theories on compact spaces) the answer is not straightforward (the rank 1 case was studied in \cite{Nieri:2013yra,Nieri:2013vba}). First of all, the free boson realization of ${\rm W}_{q,t}$ explicitly breaks the $q,t$ symmetry of the 5d theory, which is however restored in the large $N$ limit (i.e. sending the number of screening currents to infinity or going to the affine case). Secondly, 5d partition functions on $S^4\times S^1$ \cite{Kim:2012gu,Terashima:2012ra,Iqbal:2012xm}, $S^5$ \cite{Kallen:2012cs,Kallen:2012va,Imamura:2012bm,Lockhart:2012vp,Kim:2012qf,Kim:2012ava,Minahan:2013jwa}, $Y^{p,q}$ \cite{Qiu:2013pta,Qiu:2013aga} and toric Sasaki-Einstein manifolds \cite{Qiu:2014oqa} exhibit  an ${\rm SL}(3,\mathbb{Z})$ factorization property rather than just ${\rm SL}(2,\mathbb{Z})$. How this symmetry enhancement can emerge from 3d considerations is highly non-trivial and we leave this topic for future research. Here we just observe that the large $N$ limit we have just mentioned is essentially the geometric transition in open/closed topological strings \cite{Gopakumar:1998ki,Vafa:2000wi,Cachazo:2001jy,Aganagic:2002wv}. Interestingly enough, the $S^3$ and $S^5$ partition functions have been proposed \cite{Lockhart:2012vp} to give a non-perturbative definition of open and closed topological string partition functions respectively, which is consistent with the expectation that our construction can describe 5d theories on compact spaces as well.

Thirdly, the ${\rm W}_{q,t}$ modular double algebra might also be useful to study 4d supersymmetric gauge theories in all those situations where 3d theories appear as boundary conditions or interfaces \cite{Gaiotto:2008sa,Gaiotto:2008ak,Dimofte:2011ju,Dimofte:2011py,Cecotti:2011iy,Dimofte:2014ija}. In particular, S-duality domain wall in 4d $\mathcal{N}=2$ theories of class $\mathcal{S}$ \cite{Gaiotto:2009we} are realized by 3d $\mathcal{N}=2$ ${\rm U}(N)$ YM theories on $S^3_b$ \cite{Terashima:2011qi,Galakhov:2012hy,Teschner:2012em,Vartanov:2013ima,Floch:2015hwo}, and their partition functions are modular kernels of Liouville \cite{Ponsot:1999uf,Ponsot:2000mt,Teschner:2001rv} or Toda theories \cite{Floch:2015hwo}. Moreover, our construction of the ${\rm W}_{q,t}$ modular double should easily lift to the elliptic case \cite{Nieri:2015dts,Iqbal:2015fvd}. In fact, as shown in \cite{Nieri:2015dts}, 4d holomorphic blocks \cite{Nieri:2015yia} of 4d $\mathcal{N}=1$ ${\rm U}(N)$ theories on $D^2\times\mathbb{T}^2$ are captured by correlators of vertex operators and screening charges of the elliptic Virasoro algebra. Since 4d compact space partition functions (including $\mathcal{N}=1,2$ supersymmetric indexes \cite{Romelsberger:2005eg,Romelsberger:2007ec,Kinney:2005ej,Dolan:2008qi}, see also \cite{Rastelli:2014jja} for a review) can be decomposed into holomorphic blocks as much as in 3d \cite{Nieri:2015yia,Yoshida:2014qwa,Peelaers:2014ima,Chen:2014rca}, we expect that elliptic matrix models provided by Coulomb branch localization on 4d manifolds with the topology of $S^3\times S^1$, $L(r,1)\times S^1$ and $S^2\times \mathbb{T}^2$ \cite{Benini:2011nc,Razamat:2013jxa,Closset:2013sxa,Assel:2014paa,Nishioka:2014zpa,Honda:2015yha,Gadde:2015wta} can naturally be studied with the same techniques developed in this paper and \cite{Nedelin:2015mio}. This perspective might also reveal interesting connections between the elliptic ${\rm W}$ algebras and elliptic integrable systems arising in that context \cite{Gadde:2011ik,Gaiotto:2012xa,Razamat:2013qfa}.

Finally, there are mathematical aspects which deserve further investigations. For instance, our construction of the ${\rm W}_{q,t}$ modular double shares many similarities with Faddeev's $\mathcal{U}_q(\mathfrak{sl}_2)$ modular double \cite{Faddeev:1999fe} (see also \cite{Kharchev:2001rs}). In that case there are two commuting copies of $\mathcal{U}_q(\mathfrak{sl}_2)$ whose $q$-deformation parameters are related by $S\in{\rm SL}(2,\mathbb{Z})$ and a unique simultaneous $\mathcal{R}$ matrix for both copies; in our case there are two commuting ${\rm W}_{q,t}$ algebras with $q$-deformation parameters related by ${\rm SL}(2,\mathbb{Z})$ elements and a unique simultaneous screening current $\mathcal{S}$ for both copies. Interestingly enough, they are exactly the representations of the $\mathcal{U}_q(\mathfrak{sl}_2)$ modular double that are relevant in the construction of the Liouville modular kernel, which has also a 3d gauge theory interpretation as we mentioned. Therefore, we also hope that our results may help to clarify the precise relation between $\mathcal{U}_q(\mathfrak{sl}_2)$ and \mbox{$q$-Virasoro} algebras (and their modular doubles) which is still not very well understood, as well as the role of ${\rm W}_{q,t}$ algebras in ordinary 2d CFTs.

\acknowledgments
We thank Francesco Bonechi, Sara Pasquetti, Vasily Pestun and Alessandro Torrielli for discussions. The research of A.N. is supported in part by INFN and by MIUR-FIRB grant RBFR10QS5J ``String Theory and Fundamental Interactions".
The research of F.N. and  M.Z. is supported in part by Vetenskapsr\r{a}det under grant \#2014-5517, by the STINT grant and by the grant 
``Geometry and Physics" from the Knut and Alice Wallenberg foundation.

\appendix

\section{Special functions}\label{sec:specialf}
We summarize the special functions and their properties used throughout the paper, for details we refer to \cite{2003math......6164N}. The (multiple) $q$-Pochhammer symbol is defined by
\be\label{qPoch}
(x;q_1,\ldots,q_n)_\infty=\exp\left(-\sum_{k>0}\frac{x^k}{k\prod_{i=1}^n(1-q_i^k)}\right)=\prod_{k_1,\ldots,k_n\geq 0}(1-x q_1^{k_1}\cdots q_n^{k_n})~.
\ee
The last expression is valid for $|q_i|<1$, but it can be continued to other regions by means of 
\be\label{qPoch:an}
(x;q)_\infty=\frac{1}{(q^{-1}x;q^{-1})_\infty}~.
\ee
The finite $q$-Pochhammer symbol is defined by 
\be\label{nqPoch}
(x;q)_n=\frac{(x;q)_\infty}{(q^n x;q)_\infty}=\prod_{k=1}^{n-1}(1-x q^k)~, 
\ee
and satisfies 
\be\label{qrefl}
(x;q)_{-n}=(q^{-n}x;q)_n^{-1}~.
\ee
The $\Theta$ function is defined by
\be\label{Theta}
\Theta(x;q)=(x;q)_\infty(q x^{-1};q)_\infty~.
\ee
The modular properties we are interested in are
\be\label{ThetaS}
\Theta(\e^{2\pi\i X}\e^{\frac{2\pi\i\ell}{r}};\e^{2\pi\i\epsilon})\Theta(\e^{\frac{2\pi\i X}{r\epsilon-1}}\e^{-\frac{2\pi\i\ell}{r}};\e^{\frac{2\pi\i\epsilon}{r\epsilon-1}})=\e^{\frac{\i\pi}{r}\ell(r-\ell)}\e^{-\i\pi \left(B_{22}(X|1,\epsilon)+B_{22}(1+\frac{X}{r\epsilon-1}|1,\frac{\epsilon}{r\epsilon-1})\right)}~,~~~~~
\ee
for $r\in\mathbb{Z}$, $\ell\in\mathbb{Z}_r$, and
\be\label{Thetaid}
\Theta(q^{-\frac{\ell}{2}}x;q)\Theta(q^{-\frac{\ell}{2}}x^{-1};q^{-1})=(-q^{-\frac{1}{2}} x)^\ell~,
\ee
\be\label{ThetaA}
\Theta(q^{-\frac{\ell}{2}}x;q)\Theta(q^{\frac{\ell}{2}}x;q^{-1})=(-x)^{\ell+1}~,
\ee
for $\ell\in\mathbb{Z}$. Here $B_{22}(X|\ul\omega)$ is the quadratic Bernoulli polynomial
\be
B_{22}(X|\ul\omega)=\frac{1}{\omega_1\omega_2}\left(\left(X-\frac{\omega}{2}\right)^2-\frac{\omega_1^2+\omega_2^2}{12}\right)~,\quad \omega=\omega_1+\omega_2~.
\ee
The double Sine function $S_2(X|\ul\omega)$ is defined as the $\zeta$-regularized product
\be\label{S2true}
S_2(X|\ul\omega)=\prod_{n_1,n_2\geq 0}\frac{n_1\omega_1+n_2\omega_2+X}{n_1\omega_1+n_2\omega_2+\omega-X}~.
\ee
Two important properties of the double Sine function are the quasi-periodicity 
\be\label{quasiper}
\frac{S_2(X|\ul{\omega})}{S_2(b_1\omega_1+b_2\omega_2+X|\ul{\omega})}\!=\!
(-1)^{b_1 b_2}2^{b_1+b_2}\prod_{n_1=0}^{b_1-1}\sin\left(\!\pi\frac{n_1\omega_1+X}{\omega_2}\right)
\prod_{n_2=0}^{b_2-1}\sin\left(\!\pi\frac{n_2\omega_2+X}{\omega_1}\right)~,~~~~~~
\ee
and the reflection
\be\label{reflection}
S_2(X|\ul{\omega})S_2(\omega-X|\ul\omega)=1~.
\ee
For two generic complex numbers $\ul\omega$ such that ${\rm Im}(\frac{\omega_2}{\omega_1})\neq 0$, the double Sine function has the factorized expression
\be\label{S2}
S_2(X|\ul{\omega})=\e^{\frac{\i\pi}{2}B_{22}(X|\ul{\omega})}\left(\e^{\frac{2\pi\i}{\omega_1}X};\e^{2\pi\i\frac{\omega}{\omega_1}}\right)_\infty \left(\e^{\frac{2\pi\i}{\omega_2}X};\e^{2\pi\i\frac{\omega}{\omega_2}}\right)_\infty~.
\ee
The generalized double Sine function $S_{2,\ell}(X|\ul\omega)$ is \cite{Nieri:2015yia}
\be\label{genS2true}
S_{2,\ell}(X|\ul\omega)=S_2(X+\omega_1(r-[\ell]_r)|\omega,r\omega_1)S_2(X+\omega_2[\ell]_r|\omega,r\omega_2)~,
\ee
with $[\ell]_r$ the positive integer part of $\ell$ mod $r$. For two generic complex numbers $\ul\omega$ such that ${\rm Im}(\frac{\omega_2}{\omega_1})\neq 0$, we also have the factorized expression
\begin{multline}\label{genS2}
S_{2,\ell}(X|\ul\omega)=\e^{-\frac{\i\pi}{2r}[\ell]_r(r-[\ell]_r)}\e^{\frac{\i\pi}{2}\left(B_{22}(X|\omega,r\omega_1)+B_{22}(X+r\omega_2|\omega,r\omega_2)\right)}\times\\
\times(\e^{\frac{2\pi\i}{r\omega_1}(X+\omega_1\ell)};\e^{2\pi\i\frac{\omega}{r\omega_1}})_\infty (\e^{\frac{2\pi\i}{r\omega_2}(X-\omega_2\ell)};\e^{2\pi\i\frac{\omega}{r\omega_2}})_\infty~.
\end{multline}

%\bibliographystyle{JHEP}
%\bibliography{mod_double_refs}

\end{document}